\documentclass[11pt]{article}

\usepackage{amsmath,amssymb} 
 
\newtheorem{remark}{Remark}[section]	
\newtheorem{lemma}{Lemma}[section]
\newtheorem{theorem}{Theorem}[section] 

\newtheorem{proposition}{Proposition}[section]

\newtheorem{corollary}{Corollary}[section]

\newcommand{\proof}{\textsc{\small proof}\quad}
\newcommand{\qed}{\hfill \textsc{\small qed}}

\begin{document}

\title{Quantum Yang-Mills-Weyl  Dynamics in Schroedinger paradigm}
\author{Alexander  Dynin\\
\textit{\small Department of Mathematics, Ohio State University}\\
\textit{\small Columbus, OH 43210, USA}, \texttt{\small dynin@math.ohio-state.edu}}
 \maketitle

\begin{abstract}
Inspired by F. Wilczek's QCD Lite, quantum Yang-Mills-Weyl Dynamics (YMWD) describes 
quantum  interaction  between   gauge bosons  (associated with  a  simple compact  gauge Lie group $\mathbb{G}$) and larks (massless   chiral    fields  colored by   an irreducible  unitary representation of $\mathbb{G}$). Schroedinger  representation of this quantum Yang-Mills-Weyl theory  is based on a sesqui-holomorphic  operator calculus of  infinite-dimensional  operators with variational derivatives. 

The spectrum of the quantum YMWD, with  initial data in the  central euclidean ball of a radius $0<R<+\infty$,  is 
self-similar in the inverse proportion to $R$. The spectrum  is a sequence of  eigenvalues convergent to $+\infty$. The eigenvalues   have finite multiplicities  with respect to a von Neumann algebra with a regular trace.

The same holds for the quantum  self-interaction of vector Yang-Mills bosons (Theorem 4.1). Furthermore,  the  fundamental vacuum eigenvalue is a simple zero (Appendix A). Presumably, this is a solution of the existence  problem for a  quantum Yang-Mills theory that  implies  a positive spectral mass gap.
 
The rigorous mathematical theory is  non-perturbative with  a running coupling constant as the only  ad hoc   parameter.  The application of the first mathematical principles   depends essentially on the properties   of the  compact simple Lie group  $\mathbb{G}$.

\medskip
\emph{Key words}: Constructive  quantum field theory,  Yang-Mills-Weyl equations,  Variational Yang-Mills-Weyl  energy-mass operator, Gelfand nuclear triples, von Neumann algebras, Yang-Mills Millennium problem. 

 \medskip
2010 AMS Subject Clasification: 	8347, 81V05, 46.65, 35.80
 \end{abstract}

\section{Introduction}
\subsection{Preamble}

F. Wilczek's \emph{QCD Lite}    is  the Quantum Chiral Dynamics (YMWD)  of  the light $u$ and $d$  quarks   stripped of their   Lagrangian  mass terms.    Lattice simulation  of  interaction between  the massless quarks and massless gluons produces  (with no Higgs field) 99\% of  the  mass of visible  universe (cp. \textsc{\small d\"{u}rr et al}\cite{Durr} and  \textsc{\small kronfeld}\cite{Kronfeld}).

 To quote   \textsc{\small wilczek}\cite[Subsection 1.2.1]{Wilczek})
\begin{quotation}
My central points are most easily made with reference to a slight idealization of QCD 
which I call, for reasons that will be obvious, QCD Lite. It is a nonabelian gauge 
theory based on the gauge group $SU(3)$ coupled to two triplets and two anti-triplets 
of left-handed fermions, all with zero mass. Of course I have in mind that the gauge 
group represents color, and that one set of triplet and antitriplet will be identiÞed 
with the quark fields $u_{{\tiny \mbox{L}}} ,\ u_{{\tiny \mbox{R}}}$ and the other with $d_{{\tiny \mbox{L}}},\ d_{{\tiny \mbox{R}}}$. 
 
 Upon demanding renormalizability, this theory appears to contain precisely one 
parameter, the coupling g. It is, in units with $\hbar = c = 1$, a pure number. I am ignoring 
the $\theta$ parameter, which has no physical content here, since it can be absorbed into 
the definition of the quark fields. Mass terms for the gluons are forbidden by gauge 
invariance. Mass terms for the quarks are forbidden by chiral $SU(2)_{{\tiny \mbox{L}}}\times SU(2)_{{\tiny \mbox{R}}}$ flavor  symmetry.
\end{quotation} 
Mathematically, besides the coupling constants, the classical  Lagrangians of quark flavors of the standard model  differ only   in their  phenomenological  quadratic   mass terms.  After  these mass terms are discarded, as in  the QCD Lite,  the Lagrangians  become indistinguishable. This paper is   based on an  anti-normal, non-perturbative, and ghostless  second quantization  of  gauged massless  spinor \emph{larks fields}.  

 The second quantization is applied to  Noether energy-mass functional of  cutoff transverse  initial data
with compact supports.
The ensuing  quantum energy-mass spectra are   infinite sequences of eigenvalues increasing  to infinity. These spectra are  selfsimilar on the  sliding energy-mass scale.
\medskip

My  argument  is  completely different  from the Wilczek consideration. It is mathematically rigorous and, actually, holds for \emph{arbitrary} simple compact gauge Lie groups  $\mathbb{G}$  together with their unitary representations on  Hermitian \emph{color spaces}  $\mathbb{C}^n$.\footnote{Previouslly, \textsc{\small corrigan-ramond}\cite{Corrigan} used  "larks" to dub  \emph{massive} analogs  of quarks associated with $SU(N)$  for \emph{large} $N$. With E. Corrigan permission, I am using the same term for \emph{massless} analogs  of quarks.}

Representations of a compact simple Lie group are built from the fundamental ones  indexed by the vertices of the Dynkin diagram  of the group (see, e.g., \textsc{\small fulton-harris}\cite{Fulton}).
For $\mathbb{G}=\mathbf{SU}(N)$ the fundamental representations may be chosen as 
 the $N+1$  exterior powers 
$\wedge^r\mathbb{C}^N, \ r=0,1,...,N,$ of the defining representation $\varphi(g)(\vec{w})=g\vec{w}$
 \begin{eqnarray*}
 & &
\wedge^r(g)(\vec{w}_{i_1}\wedge...\wedge \vec{w}_{i_r})
:=g(\vec{w}_{i_1})\wedge...\wedge g(\vec{w}_{i_r}),\quad v_{i_j}\in\mathbb{C}^N,\\ 
 & &
\wedge^0(g):=\mbox{identity}, \quad \wedge^0\mathbb{C}^N:=\mathbb{C}. 
 \end{eqnarray*}
 The dimensions $\mbox{dim}\wedge^r\mathbb{C}^N=\mbox{dim}\wedge^{N-r}\mathbb{C}^N$ are the binomial coefficients.

 In particular, for $\mathbb{G}=\mathbf{SU}(3)$ one has the singlets $\wedge^0\mathbb{C}^3,\ \wedge^3\mathbb{C}^3$ and the triplets $\wedge^1\mathbb{C}^3,\ \wedge^2\mathbb{C}^3$. 
 Lagrangians of massless $u$- and $d$-quarks  differ  by the fundamental representations
 of the gauge group $\mathbf{SU}(3)$ and their fractional electric charges.

This example  suggests the fractional \emph{electric charges} of   larks colored by
$\wedge^r\mathbb{C}^{N}$    to  be $r/N$,  and  of the anti-larks colored by
$\wedge^r\mathbb{C}^{*N}$  to be  $-r/N$.

There was a vivid discussion  among W. Heisenberg, P. Jordan,  and W. Pauli of the corresponding "Volterra mathematics" during early years of quantum field theory . E.g.,   P. Jordan and W. Pauli  considered the eigenvalue problem  1-dimensional  variational   Schroedinger operator for  eigenfunctionals $\Psi(\phi(x))$ of massless scalar fields $\phi(x), x\in\mathbb{R}$  (\emph{Zur Quantumelectrodynamik ladungsfreier Felder,}  Zeitung  f\"{u}r Physik, \textbf{47} (1928))
\begin{equation*}
-\left(\frac{\hbar}{4\pi}\right)^2\int\!dx\, \left[
\frac{\delta^2}{\delta \phi(x)^2}
+ c^2\left(\frac{d\phi(x)}{d x}\right)^2\right]\Psi(\phi(x))=\lambda \Psi(\phi(x)).
\end{equation*}
I present a spectral theory of  variational Schroedinger operators  based on infinite-dimensional holomorphy (see e.g. \cite{Colombeau}), sesqui-holomorphic  Fock-Kree-Gelfand  nuclear triples \cite{Kree-1},  and   spectral theory in von Neumann algebras  (cp.  \cite{Grothendieck}). In particular, this is a response to
E. Witten challenge  (\cite[p. 346]{Arnold}
\begin{quotation}
Mathematically, quantum field theory involves integration, and 
elliptic operators, on infinite-dimensional spaces. Naive attempts 
to formulate such notions in infinite dimensions lead to all sorts of 
trouble. To get somewhere, one needs the very delicate constructions
considered in physics, constructions that at first sight look rather 
specialized to many mathematicians. For this reason, together with 
inherent analytical difficulties that the subject presents,
rigorous understanding has tended to lag behind development of physics.
\end{quotation}
Obviously, E. Witten gives directions to a quantum field theory in Schroedinger paradigm. This is achieved in the present paper.  The theory is non-perturbative and relates to  problematic concepts of physicists perturbative QFT as follows.
\begin{itemize}
\item  By Ladyzhenskaya ptinciple, classical solutions of \emph{non-linear} Yang-Mills(-Weyl) equations are generated by those that have compactly supported constrained initial data in $\mathbf{R}^3$.

Quantum operators are operator-valued distributions on such initial data, the  values being  linear variational operators in Gelfand-Kree nuclear triples. In  Schroedinger footsteps, quantization of  a time-preserved energy-mass functional on the nuclear space of the initial data is treated as an eigenvalue problem for the variational Yang-Mills(-Weyl) energy-mass variational  operator.

\item The energy-mass functional is  the time component of the relativistic enegy-momentum vector, and therefore is not a relativistic invariant. However, qualitatively, the theory does not depend on a choice of   relativistic coordinates.

\item The spatial cutoffs of Yang-Mills(-Weyl) energy-mass variational operators are self-similar with respect to the squared dimensionally  
transmuted coupling constant (the scaling  renormalization). 

Their self-similar spectra are bounded from below sequences of eigenvalues converging to infinity.  In agreement with the Yukawa principle, they have   positive mass gaps.

The fundamental Yang-Mills vacuum eigenvalue is a simple zero.

\item As the dimensionally transmuted coupling constant goes to zero, the Yang-Mills(-Weyl) mass gap converges to zero (asymptotic freedom).

\item The Yang-Mills(-Weyl) variational operators are local since their symbols are local polynomial functionals.

The coherent matrix elements of the variational operators corresponds to the correlation functions.They are expanded into Taylor sesqui-holomorphic series of polynomial coherent matrix elements

\end{itemize}

\emph{Acknowledgments}.  The arXiv.org site has helped greatly. Versions of my arXiv preprints originated in 2009-10  respond to  the  ensuing  criticism, especially by L. Faddeev, Cl. Taubes, and A. Neumaier. I am also grateful to M. Agranovich and M. Frasca for their moral support.

The results have been partially reported in plenary talks  at  the Moscow International Conference on the occasion of 90th Anniversary of M.I. Vishik (2012), the Moscow 3d International Conference "Theoretical Physics and its Applications" (2013), and published in Russian Journal of Mathematical Physics, Vol.  21, No. 2,  2014, pp.169-188,
and Vol.  21, No. 3,  2014, pp.326-328.

\subsection{Outline}
\begin{description}
 
 \item[Section 2.]  Classical Yang-Mills equations in the temporal gauge and the first order formalism. Ladyzhenskaya principle: by finite speed propagation,  global solutions of the Cauchy problem for  non-linear (!) hyperbolic equations are generated by solutions with compactly supported initial data. Rectification of the infinite-dimensional  manifold of constraints for  initial data with compact supports. Further reduction to  to transverse solutions of the  constraint equations. Massless spinor  lark fields as solutions of chiral Yang-Mills-Weyl equations.

  \item[Section 3.]  A review of sesqui-holomorphic Fock Kree-Gelfand nuclear triple over a complex Gelfand nuclear triple  with conjugation. 

  Following  nonlinear quantization program of  I. Segal (see, e.g.,  \textsc{\small segal}\cite{Segal-60}
\footnote{Segal conceived his program in response  to mathematical difficulties of Heisenberg non-linear quantum field theory}),   along with   Bogoliubov-Shirkov-Schwinger quantization postulate  (cp. \textsc{\small bogoliubov-shirkov} \cite[Chapter II]{Bogoliubov}),  Noether's energy-mass functional of  constrained  initial   data (supported by a ball $\mathbb{B}$ of a radius $R>0$) is  quantized as a variational operator in the sesqui-holomorphic Fock Kree-Gelfand nuclear triple over a complex Gelfand nuclear triple  with conjugation of  the   transverse  YMW initial data. 

  Calculus of functions of creation and annihilation operators with finite  degrees of  freedom (see  {\small \textsc{agarwal-wolf}\cite{Agarwal}}) is generalized to a sesqui-holomorphic calculus of variational operators in Kree-Gelfand triples. Quantized Galerkin approximation of variational operators by pseudodiffedrential operators on $\mathbb{R}^n$ as $n\rightarrow\infty$ allows not only the calculus  generalization
but also  provides a  computational techniques for solution of variational equations (cp.
{\small \textsc{dynin}\cite{Dynin-02}}).\footnote{This is a rigorous justification of {\small \textsc{gelfand-minlos}\cite{Gelfand}}.}

\item[Section 4.]   If a  semibounded from below selfadjoint operator has  resolvent relatively compact with respect   to a von Neumann algebra with a regular trace, then its spectrum is a sequence of   eigenvalues  converging to $+\infty$ and the spectral orthogonal projectors  are of the relative trace class. This entails  main Theorem \ref{pr:main}  about the spectrum of  the variational Yang-Mills-Weyl energy-mass operator  via specially constructed von Neumann algebra. Furthermore, the  spectrum is selfsimilar  in the inverse proportion to the radius  $R$ of the confining ball $\mathbb{B}(R)$.
\item[Appendix A] shows  that the fundamental  spectral value  of a cutoff quantum  Yang-Mills  energy-mass operator  is  the simple zero  eigenvalue with the vacuum eigenvector. The   direct  proof (without  von Neumann algebras)  is based on the domination over   the number operator (with simple fundamental eigenvalue) and  the standard spectral variational principle.

\item[Appendix B]  sketches the proof by Ladyzhenskaya method that the global solutions of the Cauchy problem for classical Yang-Mills-Weyl  equations exist and are   generated by  global solutions with compactly supported Cauchy data.
\end{description}

\section{Classical Yang-Mills-Weyl dynamics}
\subsection{Yang-Mills fields}
The \emph{global gauge  group}  $\mathbb{G}$ of a  Yang-Mills theory is   a  
connected semi-simple compact Lie group with the  Lie algebra $\mathfrak{g}$. 

The Lie algebra carries the \emph{adjoint representation} $\mbox{Ad}\,(g)X=gXg^{-1}, g\in\mathbb{G}, X\in \mathfrak{g}$, of the group $\mathbb{G}$ and the corresponding selfrepresentation $\mbox{ad}(X)Y=[X,Y],\ X,Y\in\mathfrak{g}$.  The adjoint  representation is orthogonal with respect to the \emph{positive  definite}  Ad-invariant  scalar  product
\begin{equation}
\label{eq:scalar}
X\cdot Y \  :=\ -\mbox{trace}(\mbox{ad}X\mbox{ad}Y), 
\end{equation}
the negative Killing form on $\mathfrak{g}$.  

There exists an orthonormal basis $\{X_k\}$ in $\mathfrak{g}$ such that\footnote{Summation over repeated indices is presumed throughout  this section.}  
\begin{equation}
\label{eq:skew}
[X_i,X_j]\ =\ c_{ijk}X_k,
\end{equation}
with the structure constants $c_{ijk}$ are skew-symmetric with respect to interchanges of all three  indices $i,j,k$.
\bigskip
Let the Minkowski space $\mathbb{M}$ be oriented and time oriented with  the Minkowski metric signature 
$(-1,1,1,1)$. In a Minkowski coordinate system $x^\mu, \mu=0,1,2,3$  the metric tensor is diagonal.
  In  the natural unit system, the time coordinate $x^0=t$. Thus  $(x^\mu)=(t,x^i),\  i=1,\ 2,\ 3$. 
  
 The \emph{local gauge  group} $\widetilde{\mathbb{G}}$ is the group of  infinitely differentiable $\mathbb{G}$-valued functions
 $g(x)$ on $\mathbb{M}$ with the pointwise group multiplication.  The  \emph{local gauge Lie algebra}  
 $\tilde{\mathfrak{g}}$ of $\mathfrak{g}$-valued functions   on 
 $\mathbb{M}$ with the pointwise Lie bracket.   
consists of  infinitely differentiable $\mathfrak{g}$-valued functions   on 
 $\mathbb{M}$ with the pointwise Lie bracket.   
 
$\widetilde{\mathbb{G}}$ acts via the pointwise adjoint action on $\widetilde{G}$   and correspondingly on  
$\mathcal{A}$, the real vector space of \emph{gauge   fields}   $A=A_\mu(x)\in\tilde{\mathfrak{g}}$. 

 \smallskip
   Gauge fields $A$ define   the \emph{covariant partial derivatives}  
   \begin{equation}\label{}
  \partial_{A\mu}X\  :=
\  \partial_\mu X- \mbox{ad}( A_\mu)X,\quad
X\in\widetilde{G}.  
\end{equation}
This definition shows that  in the natural units \emph{gauge  connections have the mass dimension} $1/[L]$. 

Any $\tilde{g}\in\widetilde{\mathbb{G}}$ defines the affine \emph{gauge transformation}  
\begin{equation}
\label{}
A_\mu\mapsto A_\mu^{\tilde{g}}:\ =\ \mbox{Ad}\,(\tilde{g})A_\mu-(\partial_\mu \tilde{g})\tilde{g}^{-1},\ A\in \mathcal{A},
\end{equation}
so that $A^{\tilde{g}_1}A^{\tilde{g}_2}=A^{\tilde{g}_1\tilde{g}_2}$.

\medskip
Yang-Mills \emph{curvature tensor} $F(A)$ is  the 
antisymmetric tensor\footnote{The dimensionless  Yang-Mills coupling  $\gamma^2_{{\tiny\mbox{YM}}}$ is set to 1}
\begin{equation}\label{}
F(A)_{\mu\nu} :=
\partial_\mu A_\nu-\partial_\nu A_\mu-[A_\mu,A_\nu].
\end{equation} 
The curvature is gauge invariant:
 \begin{equation}\label{}
\mbox{Ad}\,(g)F(A)\ =\ F(A^{g}), 
 \end{equation}
 as well as \emph{Yang-Mills Lagrangian} 
 \begin{equation}\label{}
 \label{eq:Lag}
  (1/4)F(A)^{\mu\nu}\cdot F(A)_{\mu\nu}.
  \end{equation}
 The corresponding gauge invariant  Euler-Lagrange equation is a   2nd order non-linear  partial differential equation $\partial_{A\mu}F(A)^{\mu\nu} =0$, called 
 the \emph{Yang-Mills equation} 
\begin{equation}
\label{eq:YM}
 \partial_\mu F^{\mu\nu}\ -\ [A_\mu, F^{\mu\nu}]\ =\ 0.
\end{equation}
 \emph{Yang-Mills fields} are solutions of Yang-Mills equation.

\subsection{First order formalism}
 In the temporal gauge  $A_0(t,x^k)=0$  the   2nd order Yang-Mills equation (\ref{eq:YM})  is equivalent to  the 1st order   hyperbolic system   for the time-dependent $A_j(t,x^k)$,  $E_j(t,x^k)$ on  $\mathbb{B}$ (see, e.g.,  \textsc{\small goganov-kapitanskii} \cite[Equation (1.3)]{Goganov})
\begin{equation}
\label{eq:evolution}
\partial_t A_k\  = \  E_k ,  \quad
\partial_tE_k \  = \  \partial_jF^j_k - [A_j,F^j_k],\ \quad\ F^j_k\ =\ \partial^j A_k - \partial_k A^j - [A^j,A_k].
\end{equation}
and the \emph{constraint  equations}
\begin{equation}
\label{eq:constraint}
 [A^k,E_k]    \ =\  \partial^kE_k, \quad \mbox{i.e.}\ \quad \partial_{k,A}E_k\ =\ 0.
\end{equation}
By  \textsc{\small goganov-kapitanskii} \cite{Goganov}, the evolution system is  a semilinear first order  partial differential  system  with  \emph{finite speed propagation} of the initial data, and the  initial  problem for it with  constrained initial data at $t=0$ 
\begin{equation}
\label{ }
a_k(x)\  :=
\ A(0,x_k), \ e_k(x)\  :=
\ E(0,x_k), \quad \partial^ke_k=[a_k,e,_k]
\end{equation}
is \emph{globally and uniquely solvable} in local Sobolev spaces on the whole Minkowski space  $\mathbb{M}$ (with no restrictions at the space infinity.)

This fundamental theorem has been  derived via  Ladyzhenskaya  1949  method (see  \cite{Goganov}) by a reduction to the case of initial data  on 3-dimensional balls  $\mathbb{B}=\mathbb{B}:\   |x|<R$. 

If the  constraint equations are satisfied  at $t=0$, then, in view of the evolution system, they are satisfied   for  all $t$ automatically. Thus the  \emph{1st order evolution system along with the  constraint equations for initial data is equivalent  to the 2nd order Yang-Mills system}. Moreover the constraint equations are invariant under  \emph{time independent} gauge transformations.

Consider the chain of Hilbert spaces $\mathcal{A}^s, -\infty<s<\infty,$ of (generalized) connections $a(x)$ that are completions of connections with compact supports in open balls    $\mathbb{B}$ of radius $R$  with respect to the norms 
\begin{equation}
\label{eq:SH}
 |a|_s^2 :=
\int_{\mathbb{B}}\! dx\,\big(a\cdot(1-\triangle)^sa\big) < \infty. 
\end{equation}

They define the real Gelfand nuclear triple (cp., e.g., \cite{Gelfand}) 
\begin{equation}
\label{eq:Gelf}
\mathcal{A}:\mathcal{A}\  :=
 \bigcap\mathcal{A}^s\ \subset \mathcal{A}^0\  \subset\
\mathcal{A}^*  :=
\bigcup\mathcal{A}^s,
\end{equation}
where $\mathcal{A}$ is  a nuclear countably Hilbert space with the dual $A^*$. 

Similarly we define the chain of Sobolev-Hilbert spaces $\mathcal{S}^s, -\infty<s<\infty,$ of (generalized)  scalar fields $u(x)$  on  $\mathbb{B}$ with values in $\mbox{Ad}\:\mathbb{G}$ and the Hilbert norms $|u|_s$. Let 
\begin{equation}
\label{eq:GelfS}
\mathcal{S}:\mathcal{S}\  :=
 \bigcap\mathcal{S}^s\ \subset \mathcal{S}^0\  \subset\
\mathcal{S}^*  :=
\bigcup\mathcal{S}^s
\end{equation}
be the corresponding Gelfand triple.
 
Let   $a\in\mathcal{A}^{s+3},\ s\geq 0$. Then, by Sobolev embedding theorem $a$  is  continuously $s+2$-differentiable  on  $\mathbb{B}$ and, therefore, the following  gauged  vector calculus  operators are continuous:
\begin{itemize}
  \item \emph{Gauged gradient} $\mbox{grad}^a \ :\  \mathcal{S}^{s+1}\rightarrow  \mathcal{A}^s$,
 \begin{equation}
\label{ } 
\mbox{grad}^a_{k}u \  :=\  \partial_k u - [a_k,u].
\end{equation}

 \item \emph{Gauged divergence} $\mbox{div}^a\ :\  \mathcal{A}^{s+1}\rightarrow  \mathcal{S}^s$,
 \begin{equation}
\label{eq:div}  
\mbox{div}^a\:b \  := \   \mbox{div}b - [a;b], \quad  [a;b]:= a_k b_k.
\end{equation}
 \item \emph{Gauged curl}  $ \mbox{curl}^a\  :\ \mathcal{A}^{s+1}\rightarrow  \mathcal{A}^s$,
 \begin{equation}
 \label{eq:curl}  
 \mbox{curl}^ab \  :=\ \mbox{curl}\:b - [a\stackrel{\times}{,}b], \quad  [a\stackrel{\times}{,}b]_i\  :=\ \varepsilon_{ijk}\ [a_j,b_k]. 
 \end{equation}
  \item \emph{Gauged Laplacian}  $\triangle^a:\ \mathcal{S}^{s+2}\rightarrow \mathcal{S}^s$,
  \begin{equation}
  \label{eq:Laplacian}  
\triangle^a u\  :=\ \mbox{div}^a(\mbox{grad}^au).
  \end{equation}
\end{itemize} 
 The  adjoints of the gauged   operators  are
\begin{equation}
\label{eq:adjoint}
 (\mbox{grad}^a)^*  =\ -\mbox{div}^a.
\end{equation}
\begin{lemma}
\label{pr:sur}
If  $a\in\mathcal{A}^{s+3}, s\geq 0,$ then  the  operator  $\mbox{div}^a:\mathcal{A}^{s+1}\rightarrow \mathcal{A}^s$ is surjective.
\end{lemma}
\proof
Let $\mathring{\mathcal{S}}^{s+2},\ s\geq 0,$  denote  the closure in $\mathcal{S}^{s+2}$ of the space of $a$'s with compact support in the interior of $\mathbb{B}$.
The conventional Laplacian  $\triangle^0:\ \mathring{\mathcal{S}}^{s+2}\rightarrow \mathcal{S}^s$ is an isomorphism (see, e.g., \textsc{\small agranovich}\cite{Agranovich}).

The gauged Laplacian $\triangle^a$   differs from the usual  Laplacian $\triangle^0$ by first order differential operators, and, therefore is a Fredholm operator of zero  index from $\mathring{\mathcal{S}}^{s+2}$ to 
$\mathcal{S}^s,\ s\geq 0$.

If $\triangle^au=0$ then then  $(\triangle^au)^* u =  (\mbox{grad}^a\,u)^* (\mbox{grad}^a\,u)$ , so that  $\mbox{grad}\,u=[a,u]$.
The computation
\begin{equation}
\label{ }
(1/2)\partial_k(u\cdot u)= (\partial_ku\cdot u)=[a_k,u]\cdot u= -\mbox{trace}(a_kuu-ua_ku)]=0
\end{equation}
shows that  the solutions    $u\in\mathring{\mathcal{S}}^{s+2}$ are constant. Because they vanish on the ball boundary, they vanish on the whole ball.   Since the index of the Fredholm operator $\triangle^a$ is zero,  its range  is a closed subspace with the codimension equal to the dimension  of its  null space. Thus    the operator $\mbox{div}^a\mbox{grad}^a$  is surjective, and so is $\mbox{div}^a$. \qed

\subsection{Transverse Yang-Mills fields}
 Consider the  bundles $\mathcal{C}^s, s\geq 0$ of constraint initial data   with the base  
$\mathcal{A}$ and the null spaces $\mathcal{E}^{s+1}_a$ of the operators $\mbox{div}^a:\ \mathcal{E}^{s+1}\rightarrow \mathcal{E}^{s}$ as fibers over  $a\in\mathcal{A}$. 
  
   Their intersection $\mathcal{C}$ is  a bundle of nuclear countably Hilbert spaces over the nuclear countably Hilbert base   $\mathcal{A}$.  Together with the unions of the dual spaces $\mathcal{C}^{-s}$ they form  a bundle of nuclear Gelfand triples $\mathcal{C}$ over the same base.
\begin{theorem}
\label{pr:orthogonal}
The  bundle  $\mathcal{C}$ is  smoothly \footnote{In this paper, smooth = infinitely differentiable.} trivial,  so that the total space of  $\mathcal{C}$ is smoothly isomorphic to the direct product  of  its base  $\mathcal{A}$ and the  fiber 
$\mathcal{C}_{a=0}$, the nullspace of the operator $\mbox{div}$ in $\mathcal{E}$. 
\end{theorem}
\proof
For $0\leq s\leq \infty$ consider the mapping 
\begin{equation}
\label{ }
f:\ \mathcal{A}^{s+2}\times  \mathcal{E}^{s+1}\rightarrow \mathcal{A}^{s},\quad 
f(a,e) :=
 \mbox{div}_a(e)
\end{equation}
Sobolev imbedding theorem shows  that the mapping is continuous.
Lemma \ref{pr:sur} implies  that the continuous  partial Frechet derivatives $\partial_ef(a,e)$
are  bounded  linear operators onto a fixed Hilbert space $T(s)$, the orthogonal complement of constant $a$'s.
 continuously dependent on the parameter $a \in \mathcal{A}^{s+2}$.  The  restrictions of 
 $\partial_ef(a,e)$ to the  orthogonal complements of the null spaces of  $\mbox{div}_a$ are one-to-one. By  the  implicit function theorem on Hilbert spaces (see, e.g.,  \cite{Lang}), this implies that  the explicit solutions $e=e(a)$ of the equation  $f(a,e)=0$ provide  infinitely smooth local  trivializations  of  Hilbert bundles $\mathcal{C}^s$. 

Their  intersection $\mathcal{C}=\cap\mathcal{C}^s$ is a locally trivial $C$-bundle over 
$\mathcal{A}$ with the associated locally trivial bundle of smooth $*$-orthonormal frames in the fibers.

Since $\mathcal{A}$ is a Frechet space, its smooth homothety retraction to the origin $a=0$ has a homotopy lifting to the frame space. Thus the bundle $\mathcal{C}$ is trivial, so that the total set of constraint initial data carries the global chart $\mathcal{A} \times \ \mathcal{C}_{a=0}$.\qed

\medskip
Let $\dot{\mathcal{A}}^{s}$ and  $\dot{\mathcal{E}}^{s}$ denote the nullspaces of the operator 
$\mbox{div}$ in $\mathcal{A}^s$ and $\mathcal{E}^s$.

By \textsc{\small dell'antoniio-zwanziger}\cite{Dell'Antonio}, the closures of  smooth gauge orbits in $\mathcal{H}^0 :=
\mathcal{A}^{0}$ intersect $\dot{\mathcal{A}}^{0}$. These closures  are the orbits of  the Sobolev group, the closure in  Sobolev space ${W}^{1,2}(\mathbb{B})$ of the group  of smooth gauge transformations.  (This Sobolev group  is  a topological group of continuous transformations in $\mathcal{A}^0$.)
Thus  $\mathcal{H}^0 :=
 \dot{\mathcal{A}}^0\times \dot{\mathcal{E}}^0$ is a \emph{quasi-gauge} for the orbifold of the direct product of the  parallel transports\ (i.e. every  $(a,e)\in\mathcal{H}^0$ is on an  orbit but some orbits may intersect $\mathcal{H}^0$ more than once (cp. \textsc{\small singer}\cite{Singer} and \textsc{\small narasimhan-ramadas}\cite{Narasimhan}). 

\medskip 
The Noether   \emph{Yang-Mills energy-mass  functional  of transverse YM fields} is  (see {\small \textsc{hatfield}\cite[Equation (8.9) with   $m_0=0$]{Hatfield} ) 
\begin{equation}
\label{eq:N1}
Y\ =\  \frac{1}{2}\int_{\mathbb{B}}\!dx\:
\big((\mbox{curl}\,a -  [a\stackrel{\times}{,}a])\cdot (\mbox{curl}\,a -  [a\stackrel{\times}{,}a]) +
e\cdot e \big) \geq 0,
\end{equation}
where $[a\stackrel{\times}{,}a]$ is the vector field with the $i$-th  components $\varepsilon_{ijk}[a_j,a_k]$.

\subsection{Larks}
 The real Clifford geometric  algebra $\mbox{Clif}(1,3)$ (see, e.g., \cite{Doran})    is generated by relativistic vectors $v$ of  Minkowski time-space $\mathbb{R}^{1,3}$ with its Lorentz quadratic form of signature $(+,-,-,-)$. We use the  notation 
$(t=x^0,x=x^k),\ k=1,2,3,$ for  relativistic  coordinates $x^\mu$ in a Lorentz frame with the corresponding Lorentzian  orthonormall basis  $v_0,v_1,v_2,v_3$ for $\mathbb{R}^{1,3}$ ). (In the associated vector components, $v_\nu=v_{\nu}^\mu$ so that we have a Clifford representation of Dirac matrices.)

The  complexification $\mathbb{R}^{1,3}\otimes\mathbb{C}$ is polarized as the direct sum of $P:= \mbox{Span}(v_2-iv_0,v_3-iv_1)$ and its complex conjugate $P^*:=\mbox{Span}(v_2+iv_0,v_3+iv_1)$. Both complex subspaces are isotropic and antidual with respect to the bilinear complex Lorentz form. 
\emph{Left and right Weyl spinor spaces} over $\mathbb{R}^{1,3}$ are isomorphic as Clifford modules to the 2-dimensional mutually anti-dual Grassmann algebras 
$\mbox{Gras}(P)$ and  $\mbox{Gras}(P^*)$ with  Clifford action $c(v)$ of relativistic 4-vectors $v$ on $\Psi\in\mbox{Gras}(P)$ by left exterior products with projections $v_P$ and  right  interior products  with projections $v_{P^*}$ and the adjoint action $c^*(v)$ on $\Psi^*\in\mbox{Gras}(P^*)$ (cp. \textsc{\small berline-getzler-vergne}\cite[Proposition 3.19]{Berline}).  The complex conjugation $\Psi\mapsto \Psi^*$   amounts to the change in  both time and space orientations and thus is   the  particle-antiparticle conjugation.\footnote{
"The particle-antiparticle conjugation operation [...]
must not be confused with the charge
conjugation operation [...] which, by deÞnition, flips all the charge-like quantum numbers of a 
field (electric charge, baryon number, lepton number, etc.) but leaves all the other quantum 
numbers (e.g., chirality) intact."  ( \textsc{\small akhmedov}\cite[Page 8]{Akhmedov})} 

\emph{Larks} $\Lambda$  are colored  Weyl spinors smooth fields on the  Minkowski  space, i.e. vector fields 
$\Lambda_\mu(x)\in\mathfrak{g}\otimes_{\mathbb{R}}\mathbb{C}^n$,
where $\mathbb{C}^n$ carries a unitary representation of  $\mathbf{G}$ (and then of $\mathfrak{g}$). \emph{The actions
of  $\mathbf{G}$ and  $\mathfrak{g}$ on $\mathbb{C}^n$ shall  be  denoted as $z\mapsto gz$ and  $z\mapsto Az$.}
 
 Yukawa   classical  self-interaction of larks is governed by the  massless gauge invariant relativistic Yang-Mills-Weyl (YMW) Lagrangian
 \begin{equation}
\label{eq:YMW}
\mathcal{L}_{{\tiny \mbox{YMW}}}(A,\Lambda) = \mathcal{L}_{{\tiny \mbox{YM}}}(A)\ +\ \mathcal{L}_{{\tiny \mbox{W}}}(A,\Lambda),
\end{equation}
the sum of the Yang-Mills Lagrangian $\mathcal{L}_{{\tiny \mbox{YM}}}(A)$(\ref{eq:Lag}) and the  gauged  Dirac Lagragian $\mathcal{L}_{{\tiny \mbox{W}}}(A,\Lambda):=(1/2)\Lambda^*c(v^\nu)\partial_{\nu A}\Lambda$.

In the first order formalism and the global temporal gauge $A_{00}(t,X)=0$  the Euler-Lagrange YMW equations 
for  time-dependent fields  on the initial euclidean space $\mathbb{R}^3$ 
split into the coupled Yang-Mills and gauged Dirac evolution systems 
 \begin{equation}
\label{eq:evolution}
\partial_tA_k=E_k,\ \partial_tE_k=\epsilon^{ijk}\partial_{iA}F_{jk}+\Lambda^*c(v_k)A^k\Lambda,\ 
\partial_t \Lambda=c(v^k)\partial_{kA}\Lambda,
\end{equation}
with the initial data
\begin{equation}
a(x^k)\equiv A(0,x^k),\quad e(x^k)\equiv E(0,x^k),\quad  \lambda(x^k)\equiv \Lambda(0,x^k),
\end{equation}
subject to  the coupled constraint equations (cp. \textsc{\small choquet-bruhat, christodoulou}\cite[Equation (4.4)]{Choquet} and \textsc{\small schwarz-\'{s}niatycki}\cite[Equation (1.5)]{Schwarz})
\begin{equation}
\label{}
\partial_{ka}e^k\ =\ (\lambda^*a\lambda)\cdot a.
\end{equation} 
Furthermore, by \textsc{\small choquet-bruhat, christodoulou}\cite{Choquet} and \textsc{\small schwarz-\'{s}niatycki}\cite{Schwarz}, the Cauchy problem  for evolution  YMW equations are well posed with the initial data in local Sobolev spaces   $\mathcal{W}^s_{\tiny{\mbox{loc}}}(\mathbb{R}^3)$, i.e. has the  unique solutions $a(t),\ e(t),\ \lambda(t) \in\mathcal{W}^s_{\tiny{\mbox{loc}}}(\mathbb{R}^{1,3})$, continuous with respect to the    $\mathcal{W}^s_{\tiny{\mbox{loc}}}(\mathbb{R}^3)$-norms.\footnote{The papers \cite{Choquet} and \cite{Schwarz} use different gauges, but any gauge is locally equivalent to the temporal gauge (see \textsc{\small{segal}\cite{Segal-79}}).}
In particular, the local initial   data at $t=0$ parametrize global larks fields (with no restrictions at the spatial 
infinity). Actually, there is no blow-up. See the Appendix  B for a sketch of the proof.

\medskip

The Noether    Weyl energy-mass  functional 
\begin{equation}
\label{eq:N2}
{W}(\lambda) \ =\  \frac{1}{2}\int_{\mathbb{B}}\! dx\: \lambda^*  c(v_k))i(\partial_{a}^k)\lambda.
\end{equation}
The functional  $W(\lambda)$ is  real valued (cp. \textsc{weinberg}\cite[Section 7.5, page 323]{Weinberg}) since
\begin{equation*}
\lambda^*  c(v_k) i\partial_a^k\lambda+i(\lambda^* \partial_a^k\lambda)^*\ =\ i\partial_a^k(i\lambda^*  c(v_k) i
\lambda).
\end{equation*}

The Noether  colored energy-mass functional $M(\lambda)$  of   larks  is the sum of the gauge energy--mass functional
$Y(a,e)$ and   \emph{real} gauged Weyl energy-mass  functional 
\begin{equation}
\label{eq:gW}
W_a(\lambda) \ =\  \frac{1}{2}\int_{\mathbb{B}}\! dx\: \lambda^*  c(v_k) i\partial_a^k\lambda.
\end{equation}

\section{Operator calculus}
\subsection{Review of Kree-Gelfand  triples}
Consider  a   Gelfand    triple  of densely imbedded complex topological  spaces  with conjugation (see, e.g., \textsc{\small gelfand-vilenkin}\cite{Gelfand-1})
\begin{equation}
\label{eq:Gelfand}
\mathcal{H}\ \subset\ \mathcal{H}^0\ \subset\ \mathcal{H}^*,
\end{equation} 
where 
\begin{itemize}
\item The Frechet space $\mathcal{H}$, with a Hermitian  conjugation $\zeta\mapsto \zeta^*$
of elements $\zeta$, is a nuclear countably Hilbert space, i.e.  the topological intersection (the inverse limit) of a countable nested family of Hilbert spaces $\bigcap \mathcal{H}^n \,\ n\geq 0, \mathcal{H}^{n+1}\subset\mathcal{H}^n$,  where the imbeddings are nuclear mappings with  dense  ranges  (see e.g. \textsc{\small  \textsc{gelfand-vilenkin}\cite{Gelfand}}). The  imbeddings  commute with the conjugation.
 
  \item The space $\mathcal{H}^0$ is a Hilbert space and  with   the Hermitian sesqui-linear form  $\zeta^*w$ (the notation is  bracketless as, e.g.,  in \textsc{\small berezin}\cite{Berezin-65}).

    \item The  space $\mathcal{H}^*$ of elements $\zeta^*$ is the anti-dual space of $\mathcal{H}$ of 
 continuous  anti-linear functionals  $\zeta^*w$ on  $\mathcal{H}^n$, i.e the topological union (the direct limit) of the antidual Hilbert spaces $\mathcal{H}^{n*}\subset\mathcal{H}^{n+1}$ with the induced Hermitian  conjugation.
 
 \end{itemize}
\emph{Kree-Gelfand  nuclear triple} 
 ({\small \textsc{kree} \cite{Kree-1} and \cite{Kree-2})  is a sesqui-holomorphic second   quantization of the Gelfand triple $\mathcal{H}$, 
\begin{equation}
 \mathcal{K}
\ \subset\ \mathcal{K}^0
\ \subset\ \mathcal{K}^*,
\end{equation}
where 
\begin{itemize}
 \item  The nuclear space $\mathcal{K}^*$ is the locally convex   space of  entire holomorphic functionals 
 $\Psi(\zeta)$ on $\mathcal{H}$ with the topology of compact convergence.\footnote{An entire holomorphic functional on a locally convex space is a \emph{continuous functional}  that is entire on any 
 complex line in the space  (see e.g.  {\small \textsc{colombeau}\cite{Colombeau}}).}
  
  \item The space $\mathcal{K}^0$ is  the   Hilbert space  of square integrable entire holomorphic functionals  on $\mathcal{H}^*$ with respect to  the  Minlos-Gauss probability measure 
  $d \zeta^*d \zeta\:e^{-\zeta^*\zeta}$ (see, e.g., \textsc{\small gelfand-vilenkin}\cite{Gelfand}).   
   \begin{equation}
\label{eq:form}
 \langle\ \Psi^*\ |\ \Phi\ \rangle\  :=
\ \int \!d \zeta^*d \zeta\:e^{-\zeta^*\zeta}\: \Psi(\zeta)\Phi(\zeta^*), \quad  \Psi^*, \Phi\in \mathcal{K},
 \end{equation}
where  $\Psi(\zeta)$ is the complex conjugate of $\Psi(\zeta^*)$. 

The notation 
is ambiguous because there is no Lebesgue measure $d \zeta^*d \zeta$ on the infinite-dimensional space
$\mathcal{H}^*$. However  the integral is the convergent limit of finite-dimensional integrals over $p(\mathcal{H}^*)$, where $p:\  \mathcal{H}^*\rightarrow\mathcal{H}$ are selfadjoint (aka orthogonal)   projectors of a finite rank $r(p)$
\begin{displaymath}
\ (2\pi i)^{-r(p)}\int \! ( d^{r(p)}\zeta)^*(d^{r(p)}\zeta) e^{-((p \zeta)^*(p \zeta)})\:\Psi^*(p \zeta)\Phi(p \zeta^*), 
\end{displaymath}
as the projectors $p$ strongly converge in $\mathcal{H}$  to the identity operator (cp. \textsc{\small berezin}\cite[Chapter 1, Section 2]{Berezin-65}).

Functionals $\Psi^*(p \zeta),\ \Phi(p \zeta^*)$ are cylindrical functionals of the rank of  $p$.
 They form dense subspaces in $\mathcal{H}$ and $\mathcal{H}^*$.
  \item     The nuclear  countably Hilbert space  $\mathcal{K}$  is  the  space of   entire holomorphic  functionals  $\Psi(\zeta^*)$ on $\mathcal{H}^*$ of the first order exponential growth on any 
 $\mathcal{H}^{n}*$.
  
 Actually, $\mathcal{K}$  is the space of continuous  anti-linear functionals $\Phi$ on $\mathcal{K}^*$,  in Dirac bra-ket notation inherited from (\ref{eq:form})
 \begin{equation}
\Phi(\Psi^*\ ):=\ \langle\ \Psi^*\ |\ \Phi\ \rangle\ =\ :\Psi(\Phi)
 \end{equation}
 
(By its  nuclearity,  $\mathcal{K}$ is reflexive).
     \end{itemize} 

\emph{Coherent states} $e^z(\zeta^*):=e^{\zeta^*z},\ z\in\mathcal{H}$ are in $\mathcal{K}$ and  have the following well known basic property\footnote{For starters they are straightforward on cylindrical states and then, by strong limits, are extended to all states  (cp. e.g. \cite{Dynin-02}} 
\begin{itemize}

\item The linear span of  $e^z,\ z\in\mathcal{H},$ is dense in $\mathcal{K}$, and, therefore, in $\mathcal{K}^0$ and
$\mathcal{K}^*$.

\item $ \langle\ e^z\ | \ e^w\ \rangle\ =\ e^{z^*w}$.

\item The   \emph{coherent Fourier transform}\footnote{aka, the Borel transform,  cp. \textsc{\small colombeau}\cite[Chapter 7]{Colombeau})} of $\Psi^*(z)$
\begin{equation}
   \label{eq:basis}
\Psi^*(\zeta)\mapsto\Psi(z)\  :=  e^{-z^*z}\langle\ \Psi^*\ |\ e^z\ \rangle,\quad
\Phi(\zeta^*)\mapsto\Phi(z^*)\  :=   e^{-z^*z}\langle\ e^{z^*} |\Phi\ \rangle  
   \end{equation}  
is a topological automorphism of $\mathcal{K}^*$ and  $\mathcal{K}$, unitary  on  $\mathcal{K}^0$. 
\item
The coherent  Fourier transform intertwines the operators of the directional differentiation $\partial_z$ and multiplication with  $\langle\ \zeta\ |\ z\ \rangle$ on  $\mathcal{K}^*$, and the operators of the directional differentiation $\partial_z^*$ and multiplication with  $\langle\ z\ |\ \zeta^*\ \rangle$ on  $\mathcal{K}$.
\end{itemize}

\medskip
By Grothendieck  kernel theory, the nuclearity of the  Kree-Gelfand triples implies that the locally convex vector space   of continuous linear operators  $\mathcal{K}\rightarrow\mathcal{K}^*$ is topologically isomorphic to the complete sesqui-linear  tensor products $\mathcal{K}^*\otimes\: \mathcal{K}^*$
over $^*\mathcal{H}\times \mathcal{H}$   of the pairs $(z^*,w)$ subject  to 
the Hermitian conjugation $(z^*,w)^* :=\ (z,w^*)$.

The corresponding coherent functionals are 
\begin{equation}
\label{eq:biex}
e^{(z^*,w)}\big((\zeta^*,\eta)^*\big)\ =\ e^{z^*\zeta+\eta^*w}.
\end{equation}

\subsection{Variational operators}
The spaces    $\mathcal{H}$ and $\mathcal{H}^*$ have the   linear  representation   by continuous linear transformations  into  the operators of  \emph{creation and annihilation} continuous operators of multiplication and complex directional differentiation in $\mathcal{K}$ and in $\mathcal{K}^*$
\begin{eqnarray}
& &
\big(\hat{z}\Phi\big)(\zeta^*)  :=  (\zeta^*z)\Phi(\zeta^*),\quad
\big(\widehat{z^*}\Phi\big)(\zeta^*)   := \big(\partial_{z^*}\Phi\big)(\zeta^*),
 \\ 
& &
\big(\widehat{z^*}\Psi\big)(\zeta)  :=  (z^*\zeta)\Psi(\zeta),\quad
\big(\hat{z}\Psi\big)(\zeta)   := \big(\partial_{z}\Psi\big)(\zeta),
\end{eqnarray} 
 such that
\begin{itemize}
\item The adjoint of a creation operator is the annihilation operator
\begin{equation}
\hat{z}^\dag\ =\ \widehat{z^*}\end{equation}

 \item  Bosonic commutation  relations  hold
 \begin{equation}
\label{eq:CCR}
[\widehat{z^*},\hat{w}]\ =\ z^*w;
\end{equation}
\item The coherent states  $e^{z},\ z\in \mathcal{H}$  are the eigenstates of the annihilation operators
\begin{equation}
\label{ }
\widehat{z^*} e^w\ =\ (z^*w)e^w,\quad \hat{z} e^{w^*}\ =\ (w^*z)e^{w}.
\end{equation}  
\end{itemize}

Creators and annihilators generate  strongly continuous abelian operator groups in $\mathcal{K}$ parametrized by $z$ and $z^*$:
\begin{equation}
e^{\hat{z}}\Psi(\zeta^*)\ =\ e^{\zeta^*z}\ \Psi(\zeta^*),\quad 
e^{\widehat{z^*}}\Psi(\zeta)\ =\  \Psi(\zeta^*+z^*).
\end{equation}
Thus the  operator products 
$e^{\hat{z}}e^{\widehat{w^*}}$  and $e^{\widehat{w^*}}e^{\hat{z}}$ 
are invertible continuous operators in $\mathcal{K}$.

By Baker-Campbell-Hausdorff commutator formula and the canonical commutation relations (\ref{eq:CCR}),
 \begin{equation}
\label{eq:BCH}
e^{\hat{z}}e^{\widehat{w^*}}\ =\ e^{\hat{z}+\widehat{w^*}}
e^{z^*w/2}, \quad  e^{\widehat{w^*}}e^{\hat{z}}\ =\ 
e^{\hat{z}+\widehat{w^*}}e^{-z^*w/2}.
\end{equation}
Therefore  the operator $e^{\hat{z}+\widehat{w^*}}$ is also continuous and invertible in  $\mathcal{K}$.

\medskip
As in \textsc{\small agarwal-wolf}\cite{Agarwal} in finite dimensions\footnote{The paper   \cite{Agarwal} is a formal operator algebra in terms of phase space c-equivalents of operator functions.  The 
present papers provides, in particular, a rigorous functional analysis content (cp. \textsc{\small dynin} \cite{Dynin-02} for a preliminary version).}, one quantizes the sesqui-linear  coherent Fourier transform of $M\in\mathcal{K}^*\otimes\mathcal{K}^*$
\begin{equation}
\label{ }
M(z^*,w)\ =\ \langle\ M(\zeta^*,\eta)\ |\ |(z^*,w) \rangle(\zeta^*,\eta)\rangle
\end{equation}
 as   continuous
  \emph{normal, Weyl,   anti-normal variational  operators} from 
  $\mathcal{K}$ to $ \mathcal{K}^*$
  defined by their \emph{coherent matrix elements} 
\begin{eqnarray}
\label{eq:n}
 \langle\ z\ |\ \widehat{M}_\nu\ |\ w\ \rangle\ 
&  := & 
 \big\langle \ M_\nu(\zeta^*,\eta)\  \big|\   \langle\  z |\ e^{\hat{z}}e^{\widehat{w^*}}\ |\ w\ \rangle(\zeta^*,\eta)\ \big\rangle,
\\
\label{eq:w}
\langle\ z \ |\ \widehat{M}_\varpi(\zeta^*,\eta) \ |\ w\ \rangle\ &  := & \ \big\langle \ M_\varpi(\zeta^*,\eta)\  \big|\   \langle\  z\  |\ e^{\hat{z}+\widehat{w^*}}\ |\ w\ \rangle(\zeta^*,\eta)\ \big\rangle,\\
\label{eq:anti}
 \langle\ z\ |\ \widehat{M}_\alpha\ |\ w\ \rangle\ &  := & \big\langle \ M_\alpha(\zeta^*,\eta)\ \big|\  \langle\ z\  |\ e^{\hat{z^*}}e^{\widehat{w}}\ |\ w\ \rangle(\zeta^*,\eta)\ \big\rangle.
\end{eqnarray}
The functionals $M_\nu\ M_\varpi,\ M_\alpha$ are \emph{normal, Weyl, and anti-normal co-kernels} of $\widehat{M}_\nu,\  \widehat{M}_\varpi,\  \widehat{M}_\alpha$.
\begin{proposition}
\label{pr:norm}
Any $M\in\mathcal{K}^*\otimes\mathcal{K}^*$ is the normal co-kernel  of  a unique continuous  linear operator $Q:  \mathcal{K}\rightarrow \mathcal{K}^*$.

Any continuous  linear operator $Q:  \mathcal{K}\rightarrow \mathcal{K}^*$ has a unique  normal co-kernel  $M_\nu^Q(z^*,z)=\langle z\ |Q|\ w\ \rangle$.\footnote{Cp.\textsc{\small berezin}\cite[Equation (1.7)]{Berezin-72}}.
 \end{proposition}
\proof  
The coherent kernel of the operator $e^{\hat{z}}e^{\widehat{w^*}}:\mathcal{K}\rightarrow\mathcal{K}$
\begin{equation}
\langle\ e^{z^*}\ | \ e^{\hat{z}}e^{\widehat{w^*}}\ |\ e^w\ \rangle\ \ =\\
\langle\ e^{z^*\zeta}\ | e^{\zeta^*z+w^*\eta}e^{w^*w}\ \rangle\ =\ 
e^{z^*\zeta+\eta^*w}e^{z^*z+w^*w}, 
\end{equation}
i.e. the co-kernel of  $e^{\hat{z}}e^{\widehat{w^*}}$ is $e^{z^*\zeta+\eta^*w}$, the coherent state (\ref{eq:biex}).

Therefore, by (\ref{eq:n}), any $M(\zeta^*,\eta)\in\mathcal{K}^*\otimes\mathcal{K}^*$ is a normal co-kernel of  a continuous linear operator $\mathcal{K}\rightarrow\mathcal{K}^*$, and vice versa any  continuous linear operator 
$\mathcal{K}\rightarrow\mathcal{K}^*$ has a (unique) normal co-kernel $M_\nu(\zeta^*,\eta)\in\mathcal{K}^*\otimes\mathcal{K}^*$. \qed
\begin{corollary} Any continuous operator $Q:  \mathcal{K}\rightarrow \mathcal{K}^*$ has unique Weyl and anti-normal co-kernels $M_\varpi^Q(z^*,z),\ M_\alpha^Q(z^*,z)$.
\end{corollary}
\proof
$e^{\hat{\theta}}e^{\widehat{w^*}},\ e^{\hat{z}+\widehat{w^*}}$
 are invertible in $\mathcal{K}$. \qed

\begin{corollary}
 Any continuous operator $Q:  \mathcal{K}\rightarrow \mathcal{K}^*$ has a strongly  convergent 
 expansion into a power series of $\hat{z}^j\widehat{z^*}^k, j+k \geq 0$ defined by a unique
 Taylor series expansion of its  sesqui-entire Grothendieck kernel.
\end{corollary}
  $M_\nu,\ M_\varpi$ and $M_\alpha$ are the \emph{normal, Weyl} and \emph{anti-normal
co-kernels} of the corresponding operators  and, by  (\ref{eq:BCH}), belong
to $\mathcal{K}^*\otimes\mathcal{K}^*$. As sesqui-holomorphic functionals
they have convergent Taylor expansions into homogeneous (their $n$-th differentials), so that the operators 
\begin{equation}
\label{eq:Taylor}
\widehat{M}_\nu=:M_\nu(\widehat{z^*},\hat{z}), \quad \widehat{M}_\varpi=:M_\varpi(\widehat{z^*},\hat{z}), \quad \widehat{M}_\alpha=:M_\alpha(\widehat{z^*},\hat{z})
\end{equation}
are strongly converging series of the corresponding quantizations of the $n$-th differentials (cp.
Berezin generating functionals \cite{Berezin-65}).

In view of the uniqueness  of Taylor  coefficients, the sesqui-entire functionals   are uniquely  defined  by   their restrictions   to the real diagonal 
\begin{equation}
\label{ }
\Re\big(\mathcal{H}^*\times \mathcal{H}\big)\ =\ \{(z^*,z):\  z^*\  \mbox{is  the \emph{Hermitian conjugate} of z}\}.
\end{equation}
Thus  the \emph{normal symbol} of the operator $Q$
\begin{equation}
\label{eq:nsymb}
\sigma^Q_\nu(z)\  :={M}_\nu(z^*,z), \quad (z^*,z)\in
\Re\big(\mathcal{H}^*\times \mathcal{H}\big),
\end{equation}
exists and defines $Q$ uniquely. 

Similarly, the restrictions of $M_\varpi^Q(z^*,z)$ and 
 $M_\alpha^Q(z^*,z)$ to the real diagonal $\Re\big(\mathcal{H}^*\times \mathcal{H}\big)$ define the Weyl and anti-normal symbols \emph{Weyl} and \emph{anti-normal}  symbols $\sigma_\varpi^Q(z)$  and $\sigma_\alpha^Q(z)$ of 
 $Q$.
 
 The  symbols are  real analytic functionals of $(\Re z,\Im z)$ on $\mathcal{H}$ that have a unique extension to $\mathcal{H}^*\times \mathcal{H}$ in $\mathcal{K}^*\otimes \mathcal{K}$. Thus each of the symbols defines  $Q$ uniquely. Correspondingly, 
 \begin{equation}
\label{eq:symb}
\widehat{M}_\nu\ =\ \sigma_\nu(\widehat{z^*},\hat{z}),\ \widehat{M}_\varpi\ =\ \sigma_\varpi(\widehat{z^*},\hat{z}),\ 
\widehat{M}_\alpha\ =\ \sigma_\alpha(\widehat{z^*},\hat{z}),
\end{equation}
where $z^*$ is the conjugate of $z\in\mathcal{H}$.

Since the coherent Fourier transform intertwines the operators of differentiation $\partial_z$ and  $\partial_{z^*}$
with operators of multiplication with the linear forms $\zeta$ and  $\zeta^*$, the equations (\ref{eq:BCH})) the  symbols of the same operator $Q$ are related by (\ref{eq:BCH}) as 
\begin{eqnarray}
\label{eq:wn}
\sigma^Q_\nu(z) & = &  e^{-(1/2) \partial_{z^*}\partial_z}\:\sigma^Q_\varpi(z),\\
\label{eq:wa}
\sigma^Q_\varpi(z) & = &  e^{-(1/2) \partial_{z^*}\partial_z}\:\sigma^Q_\alpha(z),\\
\label{eq:nan}
\sigma^Q_\nu(z) & = &  e^{- \partial_{z^*}\partial_z}\:\sigma^Q_\alpha(z).
\end{eqnarray}
(cp. \textsc{\small agarwal-wolf}\cite[formulas (5.29), (5.30), (5.31), page 2173]{Agarwal} in a finite-dimensional case; \textsc{\small dynin} \cite{Dynin-02}) in white noise calculus.)

 The co-kernel extension  of  a \emph{real} symbol is invariant under the complex conjugation, so that 
\begin{equation}
 \label{eq:sym}
\sigma_\nu(\widehat{z^*},\hat{z})^\dag=\sigma_\nu(\widehat{z^*},\hat{z}),\ 
\sigma_\varpi(\widehat{z^*},\hat{z})^\dag=\sigma_\varpi(\widehat{z^*},\hat{z}),\ 
\sigma_\alpha(\widehat{z^*},\hat{z})^\dag=\sigma_\alpha(\widehat{z^*},\hat{z}) 
\end{equation}
are symmetric operators from $\mathcal{K}$ to $\mathcal{K}$.
 \begin{proposition}
Let  $\mathcal{H}^0=\mathcal{L}^2(\mathbb{B})$ over a ball $\mathbb{B}: \|x\|\leq R$ in a euclidean space 
$\mathbb{R}^n$, and $\mathcal{H}$ be the nuclear Frechet space, the closure of infinitely differentiable functions $z^*(x)$  with compact support  in $\mathbb{B}$.

Consider a \emph{local sesqui-polynomial} functional on $\mathcal{H}$
\begin{equation}
\sigma_\alpha(z^*,z):= \int_{\mathbb{B}}\! d^nx P(z^*(x),z(x)),
\end{equation}
where $P(z^*,z)$ is a polynomial with constant coefficients.

Then $\sigma_\alpha(\widehat{z^*},\hat{z})$ is a continuous  linear operator from $\mathcal{K}$ to $\mathcal{K}$.
 \end{proposition}
\proof
Let $\Phi(z^*)$  be the coherent Fourier transform $\langle\ \Phi\ |\ e^{\zeta^*}\ \rangle$ of $\Phi(\zeta)$.
Then the  coherent Fourier transform of $\sigma(\widehat{z^*},\hat{z})\Phi$ is
\begin{eqnarray*}
& &
\langle\  \sigma_\alpha(\widehat{z^*},\hat{z})\Phi\  |\ e^{\zeta}\ \rangle \ =\ 
\langle\  \Phi\ |\ \overline{\sigma}_\nu(\widehat{z^*},\hat{z}) e^{\zeta}\ \rangle\ =\\
& &
\langle\  \Phi\ |\ \int_{\mathbb{B}}\! d^nx \overline{P}(e^{z^*(x)\zeta(x)},e^{z(x)+\zeta(x)}\ \rangle.
\end{eqnarray*}
Since $\overline{P}(e^{z^*(x)\zeta(x)},e^{z(x)+\zeta(x)}$ is a sesqui-holomorphic function  uniformly on $\mathbb{B}$,
the coherent Fourier  transform belongs to $\mathcal{K}$, and so is $\sigma_\alpha(\widehat{z^*},\hat{z})\Phi$.\qed
\begin{corollary}
\label{Fried}
If the anti-normal symbol $\sigma_\alpha(Q)$  is a non-negative loclal polynomial functional then, by Theorem \ref{pr:Berezin}, the operator $Q$ in $\mathcal{K}$ is non-negative, and as such admits Fridriechs extension to a selfadjoint non-negative operator in  
$\mathcal{K}^0$.
\end{corollary} 

\subsection{Quantized Galerkin approximations}

A \emph{Galerkin sequence}  $p_n,\ j=1,2,...,$ is an increasing  sequence  of  selfadjoint projectors of rank $n$ from $\mathcal{H}^*$ to $\mathcal{H}$ that strongly converge to the identity operator in 
$\mathcal{H}$. The finite-dimensional projectors induce the \emph{quantized Galerkin sequence} 
\begin{equation}
\label{ }
P_n\Psi(\zeta^*)\  :=
\ \Psi(p_n\zeta^*), \quad P_n\Psi(\zeta)\  :=
\ \Psi^*(p_n \zeta)
\end{equation}
of \emph{infinite-dimensional}  projectors in the triple $\mathcal{K}$ onto cylindrical triples isomorphic to the pulled back  sesqui-entire triples over the tautological finite-dimensional  triple  
$\mathbb{C}^n\subset \mathbb{C}^n \subset \mathbb{C}^n$.

By Proposition \ref{pr:norm}, the   compressions of operators $P_nQP_n :=
 P_nQP_n$  of $Q$ are  cylindrical pseudodifferential operators with the coherent Grothendieck kernels,
\begin{equation}
\label{ }
 \langle\ z\ |\ P_nQP_n\ | \ w\ \rangle\ =\ \langle\ e^{p_nz}\ |\ Q\ | \ e^{p_nw}\ \rangle,
 \end{equation}
i.e. pullbacks from $\mathbb{C}^j$ of finite-dimensional   pseudodifferential operators of \textsc{\small agarwal-wolf}\cite{Agarwal}.

\begin{theorem}
\label{pr:cylindrical}
Operator $Q$ is the strong limit of the cylindrical pseudodifferential operators $P_nQP_n$ on  $\mathcal{K}$.
\end{theorem}
\textsc{\small proof}\ 
The matrix element  $\langle\Psi^{*}|Q|\Phi\rangle$ is a separately continuous sesquilinear form on  the Frechet space $\mathcal{K}$. By a Banach theorem (see, e.g., \cite[v.1,Theorem V.7]{Reed}), the sesquilinear form is actually continuous 
on $\mathcal{K}$. In particular,  operator $Q$ is the weak limit of $P_nQP_n$ in $\mathcal{K}$. Since   $\mathcal{K}$ is a nuclear space,  the  weak convergence implies  the strong one in  the topology of $\mathcal{K}$. 
 \qed

As $n\rightarrow \infty$, the matrix elements
 \begin{equation}
\label{eq:convergence}
 \langle\ z\ |\ P_nQP_n\ | \ w\ \rangle\ =\ 
\langle\ e^{p_nz*}\ |\  Q\ | \ e^{p_nw}\ \rangle\ 
\longrightarrow\  \langle\ z\ |\  Q\ | \ w\ \rangle,
\end{equation}
so that symbols of the cylindrical $Q_\nu$  converge  to the corresponding symbols of $Q$.
This allows to extend to variational   operators   important results of finite-dimensionalf pseudodifferential theory. In particular, we get the following extensions for sufficient  operator positivity tests of \textsc{\small berezin}
\cite[Theorem 6]{Berezin-71} for anti-normal pseudodifferential operators and of \textsc{\small howe}\cite[Theorem 3.2.1]{Howe}  for Weyl pseudodifferential operators\footnote{By Paley-Wiener theorem, Howe's symbolic calculus is unitarily equivalent to the sesqui-holomorphic one in the case of finite  dimensions.}
\begin{proposition}
 \label{pr:Toeplitz}
Any tame operator $Q:  \mathcal{K}\rightarrow \mathcal{K}^0$ is a  diagonal compression of the multiplication with its  anti-normal  symbol (cp.  \textsc{\small {berezin}}\cite[Equation (1.5)]{Berezin-72})
\begin{equation}
\label{pr:Pi}
 \Pi \sigma_\alpha^Q(z^*,z)\Pi, \quad  \Pi:\ L^2(\mathcal{H}^*) \rightarrow   \mathcal{K}^0,
\end{equation}
where $\Pi$ is the orthogonal projector  onto $\mathcal{K}^0$
from the space  of \emph{all} square integrable functionals on $\mathcal{H}^*$ with respect to the  Minlos-Gauss probability  measure.   The projector $\Pi$ has the Grothendieck kernel $e^{z^*w}$  (cp. \textsc{\small folland}\cite[Chapter 2, Section 7]{Folland}). 
\end{proposition}
\begin{theorem}
\label{pr:Berezin} 
  If the  anti-normal symbol  $\sigma_\alpha^Q$ or the Weyl symbol  $\sigma_\varpi^Q$ of  an operator $Q$ in $\mathcal{K}$ are nonnegative functionals on $\mathcal{H}$ then $Q$ is a
   non-negative and symmetric operator.
\end{theorem}

\section{Relative  compactness}
A \emph{von Neumann algebra} (vNa) $\mathcal{N}$ is an algebra of bounded operators in a (separable) complex Hilbert space $\mathcal{H}^0$ such that its second commutant  
$\big((\mathcal{N}^\prime)\big)^\prime =\mathcal{N}$. Thus $\mathcal{N}$ is an unital algebra with a conjugation $Q^* := Q^\dag$ closed  in the weak operator topology (see, e.g. \textsc{\small{takesaki}}\cite{Takesaki}).
 
The   von Neumann algebras  $\mathcal{N}$  under consideration carry  a \emph{regular trace}, i.e. faithful normal semi-finite trace $\rho$ (see e.g. \textsc{\small{takesaki}}\cite[Definition V.2.1]{Takesaki}), i.e. $\rho$ is an additive homogeneous  positive function with values in $[0,+\infty]$ on the unitary conjugacy classes of non-negative  operators $T\in\mathcal{N}$ such that
 \begin{itemize}
  \item  $\rho(T)=0$ if and only if $T=0$,
  \item $\rho$ commutes with the supremum operation,
  \item For every $S>0$ in $\mathcal{N}$ there is a  operator  such that $0<T<S$.
\end{itemize}
\textsc{examples} of regular traces: (a) the operator trace $\mbox{tr}(T)$ on the vNa of all bounded linear operators  $T$  in a Hilbert space,  and (b) the Lebesgue integral on the vNa $L^\infty(\mathbb{R}^n)$ of multiplicators with bounded measurable  functions in the Hilbert space $L^2(\mathbb{R}^n)$.

A selfadjoint operator $A$ in $\mathcal{H}^0$ is \emph{adopted} to $\mathcal{N}$ if   its resolvent $(\lambda-A)^{-1}$  is witin $\mathcal{N}$ 
 
The relative \emph{trace class} $\{Q\in\mathcal{N}:\  \rho(Q)<\infty\}$
is the two-sided  ideal weakly dense in $\mathcal{N}$. 
Its norm closure is the  two-sided  ideal of relative $\rho$-compact operators in $\mathcal{H}^0$.

 \begin{proposition}
 \label{pr:relative}
Suppose a vNa $\mathcal{N}$ has a regular  trace $\rho$.  Let   an unbounded selfadjoint operator $A$ in $\mathcal{H}^0$ be  bounded from below and adopted by $\mathcal{N}$. 

If resolvents of $A$ are $\rho$-compact, then its spectrum is   a sequence of eigenvalues  converging to infinity. 
\end{proposition}
\proof
By \small{\textsc{ovchinnikov }\cite[Theorem 3]{Ovchinnikov-2}, the non-zero  spectrum of the relatively  compact  operator adopted by a vNa with a regular  trace $\rho$ is a  sequence of eigenvalues that is either finite, or  converges to $0$. Furthermore, the orthogonal projectors to the  spectral subspaces corresponding to non-zero eigenvalues have  finite traces. \qed

There is  the following relative version of variational Glazman Lemma
(cp. \small{\textsc{berezin-shubin}\cite[Appendix 1, Lemma 3.1]{Berezin-91}).
\begin{lemma}
Let $A\geq 0$ be an essentially  selfadjoint operator on a domain $\mathcal{D}$ with the spectral  decomposition $P(\lambda),\ \lambda\in\mathbb{R},$ of the unit operator (strongly continuous from the right: $P(\lambda+0)=P(\lambda)$), so that 
\begin{equation}
\label{ }
A\ =\ \int\! dP(\lambda)\: \lambda, \quad  P(\lambda)\in \mathcal{N}.
\end{equation}
Then the (possibly infinite) left limit 
\begin{equation}
\label{ }
\rho(P(\lambda-0))\ =\ \mbox{sup}_{P\in\mathcal{N}}\{\psi^*PAP\psi<\lambda\ :\  \rho(P)<\infty,\ P\psi\in\mathcal{D},\ 
\psi^*P\psi=1\}.
\end{equation}
\end{lemma}
\proof
Let  $\lambda(t)=\mbox{sup}\{\lambda\ :\ \rho(P(\lambda ))< t\}$ for  $t > 0$.
If the operator $(1+A)^{-1}$ is relatively  compact  then, by \small{\textsc{grothendieck}\cite[Proposition 5]{Grothendieck} (proof in  \small{\textsc{ovchinnikov }\cite[Section 2]{Ovchinnikov-1}),
\begin{equation}
\label{eq:Grothendieck}
\big(1+\lambda(t-0)\big)^{-1}=\mbox{inf}_{P\in\mathcal{N}}\{ \|(1-P)(1+A)^{-1}(1-P)\|\ :\ \ \rho(P)=1+\lambda(t)\}.
\end{equation}
\qed
Let two selfadjoint operators operators $A_1$ and $A_2$ be essentially selfadjoint on a  common domain $\mathcal{D}$ where $0\leq\psi^*A_1\psi\leq\psi^*A_2\psi $. Let $P_1(\lambda)$ and 
$P_2(\lambda)$ denote their spectral decompositions of the unit operator.  The operators are \emph{adopted } with vNa  $\mathcal{N}$, if their spectral projectors belong to  $\mathcal{N}$.
\begin{proposition}
\label{pr:geq}
    If for some $\lambda$ the operator $P_1(\lambda)$  has finite $tau$-trace, then $P_2(\lambda)$
  is n the  relative trace class as well and  $\rho\big(P_1(\lambda)\big)\leq \rho\big(P_2(\lambda)\big)$. In particular, the 
  spectra of  both $A_1$ and $A_2$   in the interval $[0,\lambda)$ are finite sets.
   \end{proposition}
   \proof This is the relative version of  Corollary 1 in  \small{\textsc{berezin-shubin}\cite[Appendix 1, Section 3]{Berezin-91}) in the classical case, where it follows from classical Glazmann lemma.  The same argument is working for the relative extension. \qed

\subsection{Free fermion field}

 In terms of Cook's particle representation of a free fermion field  (cp. {\small \textsc{baez-segal-zhou}
\cite[Chapter 2, Theorem 2.2]{Baez}) over a Gelfand complex triple with conjugation 
\begin{equation}
\label{ }
\mathcal{H}_f\subset\mathcal{H}_f^0\subset\mathcal{H}_f^*,
\end{equation}
the fermionic Kree-Gelfand triple
\begin{equation}
\label{ }
\mathcal{K}_f=\bigoplus_{n=0}^\infty \mathcal{K}_f^{(n)}\
\subset \ \mathcal{K}_f^0=\bigoplus_{n=0}^\infty \mathcal{K}_f{0(n)}\
\subset \ \mathcal{K}_f^*=\bigoplus_{n=0}^\infty \mathcal{K}_f^{*(n)}\
\end{equation}
of  the  rank $n$ antisymmetric tensors (over  correspondingly $\mathcal{H}_f,\ \mathcal{H}_f^0,\ \mathcal{H}_f^*$)
carries the  \emph{differential quantization representation} $Q_C$ of (un)bounded operators $C$ in $\mathcal{H}_f^0$  essentially selfadjoint on a domain $\mathcal{H}_f$. The representation is  the direct sum  $Q_C=\bigoplus_{n\geq 0}Q_{C}^{(n)}$, where $Q_{C}^{(n)})$ are  uniquely defined by
\begin{eqnarray*}
& &
Q_{C}^{(n)}(\psi_1\wedge\ldots\wedge\psi_n)\ :=\ \bigoplus_{j>0}\psi_1\wedge\ldots\wedge\psi_{j-1}\wedge
(C\psi_j)\wedge\psi_{j+1}\wedge\ldots\wedge\psi_n,\\ 
& &
 \psi_1,\ldots,\psi_n\in\mathcal{H}_f. 
\end{eqnarray*}
Thus the spectral projectors of   selfadjoint extensions of $C^{(n)}$ belong to the vNa $W^*$ tensor product  
$\mathcal{B}(\mathcal{H}^0)\overline{\otimes}
 \mathcal{P}$, where $\mathcal{P}$ is the commutative vNa generated by the orthogonal projectors
  $P^{(n)}:\ \mathcal{K}_f^0\rightarrow \mathcal{K}_f^{0(n)}$.

Let $\mathcal{B}(\mathcal{H}_f^0)$ be the vNa of bounded linear operators in $\mathcal{H}_f^0$ and $\mathcal{P}$ be the commutative vNa generated in $\mathcal{K}^0$ by the orthogonal projectors to  $\mathcal{K}_n^0$.

The vNa   $\mathcal{P}$ is isomorphic to the vNa $l^\infty$ of bounded sequences
$\xi=\{\xi_n\}_{n=0}^\infty$ in the  Hilbert space of sequences
$\{\xi_n\}_{n=0}^\infty:\ \xi^*\xi=\sum_{n=0}^\infty|\xi_n|^2<\infty$.

Thus the $W^*$ tensor product $\mathcal{B}(\mathcal{H}_f^0)\overline{\otimes}\mathcal{P}$ carries the regular trace $\rho_{\mbox{\textsc{\tiny {W}}}}$ uniquely defined by
\begin{equation}
\rho_{\mbox{\textsc{\tiny {W}}}}(T\overline{\otimes}\{\xi_n\} :=\  \mbox{tr}(T)\Big(\sum_{n=0}^\infty\xi_n\Big).
\end{equation}
(Here $\mbox{tr}(T)$ denotes the standard operator trace on $\mathcal{B}(\mathcal{H}_f^0)$.
\begin{proposition}
\label{pr:rho}
If  $C$  is an elliptic operator in $\mathcal{H}^0$ that is semibounded   from below, essentially selfadjoint  and  with compact resolvent then the quadratic operator $Q_C$ is a semibounded   from below essentially selfadjoint operator in $\mathcal{K}^0$ and its spectrum is a sequence of eigenvalues  converging to $+\infty$.
\end{proposition}
\label{pr:adopted 1}
\proof
The essential selfadjointness of $Q_C$ has been established already.

The compactness of the resolvent of the semibounded $C$ entails the compactness of the spectral projectors $p$ of $C$ in
$\mathcal{H}_f^0$.

 Then equation(\ref{eq:n}) entails that the spectral projectors  of $Q_C^{-1}$ are relatively compact operators  in  $\mathcal{B}(\mathcal{H}_f^0)\overline{\otimes}\mathcal{P}$ because as  uniform operator limits of their relatively compact representatives $p\overline{\otimes}\{1/n\}_1^\infty$ in 
 $\mathcal{B}(\mathcal{H}_f^0)\overline{\otimes}l^\infty$. \qed
  
\medskip
\textsc{example}:\ Let $\mathcal{H}_f^0$ be the complex Hilbert space $\psi\in\mathcal{H}_f^0$ of left 
lark fields $\psi=\psi_L$ on the ball $\mathbb{B}$ and $C=W_0$. The the \emph{naked} Weyl operator (where $\psi$ denotes both left and right lark fields) 
\begin{equation}
\label{eq:0W}
W_0(\psi) \ =\  \frac{1}{2}\int_{\mathbb{B}}\! dx\:  c(v_k) i\partial^k\psi .
\end{equation}
is  essentially selfadjoint on the smooth lark fields with compact 
supports in the interior of $\mathbb{B}$. The operator $W_0$ is essentially selfadjoint and elliptic but not semibounded from below. Following \textsc{baez-segal-zhou}\cite[Page 165]{Baez} we convert  the
multiplication with the complex unit $i$ on $\mathcal{H}_f^0$  into  multiplication with  $i^\prime:=i\mbox{sign}(\psi^*W_0\psi$. The new complex unit squares to $-1$, leaves invariant the Hermitian form $\psi^*\psi$, and therefore retaining  the fermionic  commutation relations 
\begin{equation}
\label{ }
\widehat{\psi^*}\hat{\psi}+\widehat{\psi^*}\hat{\psi}=\psi^*\psi, \quad \psi^*,\psi\in\mathcal{H}_f^0.
\end{equation}
Now $\mathcal{H}_f^0$ splits into  the direct sum $\mathcal{H}_f^{0+}\oplus\mathcal{H}_f^{0-}$ of
Hilbert subspaces determined by  the $ \mbox{sign}(\psi^*\mathbf{W}_0\psi)$.

In the \emph{modified Weyl operator} $\mathbf{W}_0^\prime$ on $\mathcal{H}_f^0$ the complex unit $i$ is replaced by $i^\prime$. Now $\mathbf{W}_0^\prime$ is  nonnegative and elliptic
in $\mathcal{H}_f^0$, so that its  resolvent is  relatively compact in 
$\mathcal{B}(\mathcal{H}^0)\overline{\otimes}\mathcal{P}$. Then Propostition \ref{pr:rho} entails
\begin{proposition}
The modified energy-mass operator $Q_{\mathbf{W}_0^\prime}$ is essentially selfadjoint in
 $\mathcal{K}_f^0$ and  has the  relatively compact resolvent in 
 $\mathcal{B}(\mathcal{H}_f^0)\overline{\otimes}\mathcal{P}$.
\end{proposition}

\subsection{YM energy-mass operator}
  The \emph{anti-normal symbol} $Y(z^*,z)=Y(a,e)$ (\ref{eq:N1}) is a non-negative  local   polynomial functional on 
  $\Re(\mathcal{H}^*\times\mathcal{H}^*)$. Then, by Corollary \ref{pr:Berezin}, the  Yang-Mills energy-mass operator
   $\widehat{Y_\alpha}$ has a unique  non-negative selfadjoint  Friedrichs extension $\mathbf{Y}$ in  $\mathcal{K}^0$. 
The pointwise  product $a(x)b(x)$ of transverse fields is transverse, so that the nuclear space $\dot{\mathcal{A}}$ is closed under  the Lie bracket $a(x)b(x)-b(x)a(x)$.
Let  $\mathcal{U}$ be the  von Neumann algebra (vNa) of operators in $\mathcal{H}^0$ generated by the adjoint representation of  $\dot{\widetilde{\mathbb{G}}}$.

 The transversity condition $\mbox{div}\,a=0$ implies that  $a$ has a matrix vector potential on $\mathbb{B}(R)$ so that  $\mbox{curl}\,a$ is zero everywhere. Thus on transverse  $(a,e)$ the energy-mass functional
(\ref{eq:N1}) is reduced to
\begin{equation}
\label{eq:tr}
Y(a,e)\  =
\ (1/2)\int_{\mathbb{B}}\!d^3x\:
 \big( [a\stackrel{\times}{,}a])\cdot[a\stackrel{\times}{,}a]\ +\ e\cdot e \big).
\end{equation} 
The adjoint action  preserves both integrand terms. 
 \begin{proposition}
\label{pr:commutant}
The commutant $\mathcal{U}^\prime$ of the algebra $\mathcal{U}$ is  the 
center $\mathcal{Z}=\mathcal{U}\cap \mathcal{U}^\prime$  of both
$\mathcal{U}$ and $\mathcal{U}^\prime$ isomorphic to the vNa 
$L^\infty(\mathbb{B})$. 
\end{proposition}
\proof
The algebra $\mathcal{U}$ is generated  by operators 
\begin{equation}
\label{eq:Gr}
z(x)\mapsto \big(\mbox{Ad}\:g(x)\big)z(x), \quad z\in \mathcal{H},\ g\in\dot{\widetilde{\mathbb{G}}}.
\end{equation}
These operators  preserve zeroes   of $z$ on $\mathbb{B}$, i.e. 
\begin{equation*}
\label{ }
\{x:\  z(x)=0\}\ =\ \{x:\  (\mbox{Ad}g(x))z(x)=0\}.
\end{equation*}
Thus the Grothendieck kernel $K(x,y)$ of the transformation (\ref{eq:Gr}) is a finite order distribution on the diagonal $\{y=x\}$. Since the operator is bounded in the Hilbert space $\mathcal{H}_0^0$, it is a multiplication with a continuous $\mathbb{G}$-valued function. The function commutes with all transformations (\ref{eq:Gr}). Since  the adjoint representation of simple Lie group 
$\mathbb{G}$ is irreducible and   $\mathbb{B}$ is compact, the fundamental Schur commutator lemma  implies that $z(x)$ is a   bounded continuous scalar matrix function. 

Finally, $L^\infty(\mathbb{B})$ is the  second commutant of the $\mbox{C}^*$-algebra of continuous scalar funcions on $\mathbb{B}$. \qed

The space  $\mathcal{H}^0$ as the  Hilbert tensor product 
$L^2(\mathbb{B}0\otimes \mathbb{C}\mathfrak{g})$
(see(\ref{eq:GelfS})) carries the $W^*$ von Neumann tensor product (see \textsc{\small takesaki}\cite[Chapter IV, Definition 5.1]{Takesaki})
$\mathcal{Z}\overline{\otimes}\mathcal{B}(\mathbb{C}\mathfrak{g})$ of  the center algebra
$\mathcal{Z}$ and algebra $\mathcal{B}\big(\mathbb{C}\mathfrak{g})$ of linear transformations of the finite-dimensional  vector space $\mathbb{C}\mathfrak{g}$.

Let
\begin{equation}
\mathcal{K}(\mathbb{C}\mathfrak{g})\ \subset\ \mathcal{K}^0(\mathbb{C}\mathfrak{g})\ \subset\
\mathcal{K}^*(\mathbb{C}\mathfrak{g}) 
\end{equation}
denote the Kree-Gelfand triple over the complexification of the finite-dimensional real  vector space $\mathfrak{g}$ with the natural conjugation in $\mathbb{C}\mathfrak{g}$.  

Central operators $z\in\mathcal{Z}$ act  in $\mathcal{K}^0(\mathbb{C}\mathfrak{g})$ as $z\Psi(w^*)=\Psi(zw^*)$.

The vNa $\mathcal{Z}\overline{\otimes} \mathcal{B}\big(\mathcal{K}^0(\mathbb{C}\mathfrak{g})\big)$ in $\mathcal{K}^0$ carries  the regular trace $\rho_{\mbox{\textsc{\tiny Y}}}$ uniquely defined by  
\begin{equation}
\label{}
 \rho_{\mbox{\textsc{\tiny Y}}}(z\overline{\otimes}T)\ :=\ \int_\mathbb{B}\!dx\:z(x)\mbox{tr}(T),\quad z\in \mathcal{Z}=
 L^\infty(\mathbb{B}),\ T\in\mathcal{B}\big(\mathcal{K}^0\big(\mathbb{C}\mathfrak{g})\big).
 \end{equation} 

\begin{theorem}
\label{pr:adopted 2}
The  Yang-Mills energy-mass operator $\mathbf{Y}$  is non-negative and  essentially selfadjoint on $\mathcal{K}$. It is adopted by the  vNa $\mathcal{Z}\overline{\otimes} \mathcal{B}\big(\mathcal{K}^0(\mathbb{C}\mathfrak{g})\big)$
and has  $\rho_{\mbox{\textsc{\tiny Y}}}$-compact resolvent.
\end{theorem}
\proof
\begin{itemize}
  \item By (\ref{pr:Pi}), $\mathbf{Y}=\Pi \mbox{Y}\Pi$ is the orthogonal compression to $\mathcal{K}^0$ of the mutiplication operator with its transverse anti-normal symbol $Y$ (\ref{eq:tr}) in $L^2(\mathcal{K}^*)$. The orthogonal projector $\Pi$ has the coherent Grothendieck kernel  $e^{z^*w}$. Both transverse anti-normal symbol $Y$ and  Grothendieck kernel  $e^{z^*w}$ are conserved by 
  $1\overline{\otimes} \mathcal{B}\big(\mathcal{K}^0(\mathbb{C}\mathfrak{g})\big)$-action, but 
neither is $\mathcal{Z}$-covariant.  Thus
 \begin{equation}
\label{ }
 \mathbf{Y}\ =\ \mathbf{1}\overline{\otimes}\mathbf{Y}^\#,\quad \mathbf{Y}^\#\ :=\
 \Pi^\#Y^\#\Pi^\#\in\mathcal{B}\big(\mathcal{K}^0(\mathbb{C}\mathfrak{g})\big)
\end{equation}
where $Y^\#$ stands for the multiplication operator in $L^2\big(\mathcal{K}^*
(\mathbb{C}\mathfrak{g})\big)$ and $\Pi^\#$ is the orthogonal projection from $L^2\big(\mathcal{K}^*
(\mathbb{C}\mathfrak{g})\big)$ onto $\mathcal{K}(\mathbb{C}\mathfrak{g})$ with the coherent Grothendieck kernel $\exp(z^{\#*}z^\#)$.

\item The compression of the multiplication with non-negative $Y$ is a non-negative operator.
 
\item
By  (\ref{eq:wa}), the   Weyl symbol $\sigma^\mathbf{Y}_\varpi$
of the anti-normal Yang-Mills energy-mass operator  
$\mathbf{Y}$  is
\label{ } 
\begin{eqnarray}
& &
\big(\mathbf{1}\ +\ (1/2) \partial^2/\partial_a^2\ +\ (1/2)^2( \partial^2/\partial_a^2)^2\big)Y^\#(a,e)
\nonumber\\
& &
\label{eq:Y}\\
& &
\ +\ \big(\mathbf{1}\ +\ (1/2) \partial^2/\partial_e^2\ +\ (1/2)^2( \partial^2/\partial_e^2)^2\big)e^*e.
\nonumber
\end{eqnarray} 
The differential operator $\partial^2/\partial_a^2$  is invariant   under the orthogonal transformation from $a(x)$ to $\alpha^k_i(x)$, so that, by \textsc{\small simon}\cite[page 217]{Simon}),
 \begin{equation}
\label{ }
\partial^2/\partial_a^2Y^\#(a,e)\ =\ \int_{\mathbb{B}}\!dx\:(1/2)\partial^2/\partial(\alpha^k_i(x))^2
2\sum_k(\alpha^k_i\alpha^k_jc_{ijk })^2(x).
\end{equation}
The sqew-symmetry of $c_{ijk}$ implies that  $\sum_k\alpha^k_i\alpha^k_jc_{ijk}$  does not contain  $(\alpha^k_i)^2$. Then, by a Leibniz formula, 
\begin{eqnarray*}
& &
\partial^2/\partial(\alpha^k_i)^2\sum_k(\alpha^k_i\alpha^k_jc_{ijk })^2\ =\
2\big(\partial^2/\partial(\alpha^k_i)^2\sum_k(\alpha^k_i\alpha^k_jc_{ijk }\big)
(\sum_k(\alpha^k_i\alpha^k_jc_{ijk }\big)\\
& &
+\ 2\big(\partial/\partial\alpha^k_i\sum_k\alpha^k_i\alpha^k_jc_{ijk}\big)
\big(\partial/\partial\alpha^k_i\sum_k\alpha^k_i\alpha^k_jc_{ijk}\big)\\
& &
=\ 2\sum_{ijkl}\alpha^k_ic_{ijk}\alpha^l_jc_{ljk}(x)\ =\ 2a(x)\cdot a(x).\quad 
\mbox{(See  \textsc{\small simon}\cite[page 217]{Simon})}.
\end{eqnarray*} 
Thus
\begin{equation}
\label{eq:1}
\partial^2/\partial_a^2Y^\#(a,e)\ =\ a^*a,\ \mbox{and then}\ (\partial^2/\partial_a^2)^2Y^\#(a)\ =\ 2.
\end{equation}
Besides, 
\begin{equation}
\label{eq:2}
(\partial^2/\partial e^2)e^*e \ =\ 2, \quad \big(\partial^2/\partial_e^2\big)^2e^*e \ =\ 0.
\end{equation}
Equations (\ref{eq:Y}), (\ref{eq:1}), (\ref{eq:2}) show that  the   the Weyl symbol integrand of $\widehat{M}_\alpha$ 
is 
\begin{eqnarray}
& &
Y^\#(a,e)+(1/2)\ a^*a +3/2  \ =\ (1/2)\big([a\stackrel{\times}{,}a])\cdot[a\stackrel{\times}{,}a] +
a^*a + e^*e  + 3\big) \nonumber\\
& &
>\ (1/2)(a^*a + e^*e)\ =\ z^*z. \label{eq:>}
\end{eqnarray}

\item
From  (\ref{eq:>}) we get the inequality of the Weyl symbols integrands
\begin{equation}
\label{ }
\sigma^{\mathbf{Y}^\#}_\varpi(z^*,z)\ >\ z^*z\ -\ 1/2\ \stackrel{(\ref{eq:wn})}{=}\ 
\sigma^{\mathbf{Y}^\#}_\varpi(z^*,z).
\end{equation} 
The positive   difference  
\begin{equation*}
\ \sigma^{\mathbf{Y}^\#}_\varpi\ -\ \sigma^{\mathbf{N}^\#}_\varpi\ =\ 
(1/2)[a\stackrel{\times}{,}a])\cdot[a\stackrel{\times}{,}a] + 2
\end{equation*}
is the positive Weyl symbol integrand of a local polynomial  operator $Q$. It follows,  by Proposition  \ref{pr:Berezin}, 
 that $Q$ is  positive so that
\begin{equation}
\label{eq:geq}
 Y^\# \geq \mathbf{N}^\#,\ \mbox{so that}\ \mathbf{Y}=1\overline{\otimes} \mathbf{Y}^\#
\geq 1\overline{\otimes} \mathbf{N}^\#.
\end{equation}
\label{eq:in}
The operator $\mathbf{N}^\#$  is a non-negative elliptic operator on the finite-dimensional 
$\mathbb{C}\mathfrak{g}$. Then
$1\overline{\otimes} \mathbf{N}^\#$ has $\rho_{\textsc{\small Y}}$-compact resolvent. Now Proposition \ref{pr:geq} and  the unequality (\ref{eq:in}) establish the reslvent $\rho_{\textsc{\small Y}}$-compactness. 
of $\mathbf{Y}$.  \qed
 \end{itemize}
 
\subsection{YMW spectrum}
	
The modified   \emph{YMW energy-mass operator} in $\mathcal{K}\overline{\otimes}\mathcal{K}_f$ is the gauged sum 
\begin{equation}
\label{ }
\mathbf{M}^\prime\ :=\ \mathbf{Y}\overline{\otimes}\mathbf{1}\ \ + \  \mathbf{1}\overline{\otimes}
Q_{C^\prime},\quad C^\prime:=i^\prime(c(v_k) i\partial_a^k.
\end{equation}

By Proposition \ref{pr:Berezin}, $\mathbf{M}^\prime$ has a
unique semibounded  selfadjoint shifted Friedrichs extension  $\mathbf{M}^\prime$  in 
$\mathcal{K}_f^0$ (in the transfered notation).

\begin{theorem}
\label{pr:main}
The spectrum of the modified YMW energy-mass operator $\mathbf{M}^\prime$  is  a  sequence of  eigenvalues converging to  $+\infty$. Furthermore, the orthogonal projectors to corresponding eigenspaces have finite $\rho_{\small Y}\!\rho_{\small W}$-traces.
\end{theorem}
\proof  
Operator $\mathbf{M}^\prime$ is an operator in $\big(\mathcal{Z}\overline{\otimes}\mathcal{K}^0(\mathbb{C}\mathfrak{g})\big)\overline{\otimes}\mathcal{K}_f^0$ essentially selfadjoint  in $\mathcal{K}(\mathbb{C}\mathfrak{g})\overline{\otimes}\mathcal{K}_f$. 

Since 
\begin{equation}
\label{eq:Mbar}
\mathbf{M}^\prime\ =\ \mathbf{Y}^\#\overline{\otimes}\mathbf{1}\ +\ (\mathbf{1}\overline{\otimes}\mathbf{W}^\prime)\overline{\otimes}(\oplus_{n=0}^\infty P_n),
\end{equation}
 the modified YMW energy-mass operator  is adopted, by Theorem \ref{pr:adopted 2} and Proposition \ref{pr:adopted 1},  with the vNa $\mathcal{Z}\overline{\otimes}\mathcal{B}\big(\mathcal{K}^0(\mathbb{C}\mathfrak{g}\overline{\otimes}\mathcal{H}_f^0)\big)\overline{\otimes}\mathcal{P}$. 

The selfadjoint number operator $\mathbf{N}^\#$ in $\mathcal{K}^0(\mathbb{C}\mathfrak{g})$ on the finite-dimensional space $\mathbb{C}\mathfrak{g}$ is unitarily equivalent to the nonnegative elliptic harmonic oscillator in
$L^2(\mathbb{C}\mathfrak{g})$  shifted by $(1/2)\mathbf{1}$.
Thus $\mathbf{N}^\#$ has compact resolvent in $\mathcal{B}\big(\mathcal{K}^0(\mathbb{C}\mathfrak{g})\big)$.

The nonnegative selfadjoint  operator $ \mathbf{W}^\prime_0$ on  $\mathbb{B}$ with the homogeneous Dirichlet conditions is  elliptic (see e.g. {\small \textsc{agranovich}\cite{Agranovich}).
and, therefore has compact resolvent as well.

The resolvent $(\mathbf{1} + \mathbf{N}^\#\oplus\mathbf{W}^\prime_0)^{-1}$, as  the direct sum of the resolvents $(\mathbf{1} + \mathbf{N}^\#)^{-1}\oplus(\mathbf{1}+\mathbf{W}^\prime_0)^{-1}$, is a parametrix for
the operator 
\begin{equation}
\label{ }
\mathbf{N}^\#\oplus\mathbf{W}^\prime=\mathbf{N}^\#\oplus\big(\mathbf{W}^\prime_0\ +\
i^\prime Q_{C^\prime\overline{\otimes}\lambda}\big)
\end{equation}
since the operator products   $a_k(\mathbf{1} + \mathbf{N}^\#)^{-1}$ are compact operators in
 $\mathcal{K}^0(\mathbb{C}\mathfrak{g})$, the operator product 
 $(\mathbf{1}+\mathbf{W}^\prime_0)^{-1}\lambda$  is a compact operator in $\mathcal{H}_f^0$, and  tensor products of  compact operators are compact.\footnote{It is essential that the pseudodifferential operators of   negative orders  on  finite-dimensional domains are compact (see {\small \textsc{shubin}\cite{Shubin}}).} and  {\small \textsc{agranovich}\cite{Agranovich}).}

 Thus the operator $\mathbf{N}^\#\oplus\mathbf{W}^\prime$ has compact resolvent in
 $\mathcal{B}(\mathbb{C}\mathfrak{g})\oplus\mathcal{H}^0_f)$.

Now Equations (\ref{eq:geq}) and (\ref{eq:Mbar}) imply the operator inequality
\begin{equation}
\label{ }
 \mathbf{Y}^\#+\mathbf{W}^\prime\ \geq\ \mathbf{N}^\#\ +\ \mathbf{W}^\prime.
 \end{equation}
Therefore, by Proposition \ref{pr:geq},
 the spectrum of the modified YMW energy-mass operator $\mathbf{M}^\prime$  is  a  sequence of  eigenvalues converging to  $+\infty$.
 \qed

The  restriction  of the energy-mass functional $M^\prime(a,e)$  to a  ball  $\mathbb{B}=\mathbb{B}(R)$ 
the radius  $R$ is associated   with the following spectral renormalization:
\begin{proposition}
\label{pr:ss}
The spectra of cut-off quantum Yang-Mills  energy-mass
operators  are  selfsimilar in the   inverse  proportion to  the radius  $R$.
 \end{proposition}
\textsc{\small proof}\ 
The scaling transformation
\begin{equation}
\label{ }
\check{x} := x/R, \quad \check{a} := a/R, \quad \check{e} := e/R^2
\end{equation}
converts the energy-mass
functional $\mathcal{M}^\prime$ over  $\mathbb{B}$
 into the scaled energy-mass functional over $\mathbb{B}(1)$.
 
  Moreover, the scaling transformation is  canonical: it preserves the Hermitian form 
   $z_1^*z_2$  so that the quantum canonical relations are conserved under the scaling. \qed 

Reclaim the dimensionless  coupling constant $\gamma$ in the (self)interactions terms. Since the scaling transformation converts 
$\gamma$ into $\gamma\sqrt{R}$, the (self)interaction vanishes  as $R\rightarrow 0$. This means the   \emph{asymptotic freedom}!

\section{Appendix A}
The  7th Millennium  problem of Clay Mathematics Institute 
\begin{quotation}
\textsl{Prove that for any compact simple global gauge group, a nontrivial  quantum Yang-Mills theory exists on the four-dimensional Minkowski spacetime and has a positive mass gap} (cp. \textsc{witten}\cite[p. 24]{Witten}).\footnote{The official formulation of the Millennium Yang-Mills problem (cp. \textsc{jaffe-witten}\cite{Clay}) is looking for   a quantum Yang-Mills theory as deep as   the axiomatic  relativistic  quantum field theory. However,  even modified  Wightman  axioms  (see, e.g., \textsc{bogoliubov} et al \cite[chapter 10]{Bogoliubov} are in a serious conflict  with the simplest cases of Gupta-Bleuler  theory of quantum  electromagnetic fields, as well as with common local renormalizable gauges (see, e.g.  \textsc{strocchi}\cite[Chapter 6 and Appendix A.2]{Strocchi}).}\end{quotation} 
By \textsc{goganov-Kapitanskii}\cite{Goganov}, smooth global solutions of Yang-Mills equations (in the Schwinger 1st order formalism) with no restrictions for Cauchy data at the spatial infinity are uniquely defined by  those with  supports  in central balls $\mathbb{B}=\mathbb{B}(R)$ of arbitrary radius $R>0$. 

The quartic Yang-Mills energy-mass  functional  (\ref{eq:tr})
  on the transversal nuclear Gelfand triple
$\mathcal{H}$   over $\mathbb{B}$ 
\begin{equation}
\label{eq:tr}
Y(a,e)\  =
\ (1/2)\int_{\mathbb{B}}\!d^3x\:
 \big( [a\stackrel{\times}{,}a])\cdot[a\stackrel{\times}{,}a]\ +\ e\cdot e \big).
\end{equation} 
is  the \emph{anti-normal symbol} of 
the  quantum Yang-Mills energy-mass 
operator $\mathbf{Y}\ :=\ \widehat{M}_\alpha$ in the Gelfand-Kree triple $\mathcal{K}$.
It has a unique nonnegative selfadjoint Friedrichs extension  $\mathbf{Y}$  in the transversal
$\mathcal{K}^0$ (the notation does not change).

\begin{theorem}
The fundamental  spectral value  of the quantum energy-mass operator  $\mathbf{Y}$ is  the simple zero  eigenvalue with the vacuum eigenvector.
\end{theorem}
\proof  
The  tame quadratic operator 
\begin{equation}
\label{ }
\mathbf{N}\:=\ :\ \partial_{z^*}\partial_{z}:\ \mathcal{K}\ \rightarrow \mathcal{K},\quad \sigma^\mathbf{N}_\nu\ =\ z^*z,
\end{equation}
is the \emph{number operator}.

Power  states  $z^{*n},\ n=0,\ 1,...,$ form an  orthogonal set in  $\mathcal{K}^0$ and
are eigenvectors of} $\mathbf{N}$: 
\begin{equation}
\label{eq:spec}
\mathbf{N}(z^{*n})\ =\ nz^{*n}, n=0,\ 1,\ 2,\ ...\ ,
\end{equation}
with the dense  linear span  in $\mathcal{K}^0$.\footnote{The  monomials are cylindrical states in the Bargman-Segal space $\mathcal{K}^0$ of co-rank $1$. Then the  Lemma holds  because it is true in the Bargman-Segal space  on the complex line  $\mathbb{C}$.}
Therefore   the number operator has the unique  self-adjoint Friedrichs extension $\mathbf{Y}$ 
the  notation is preserved) with the  spectrum  consisting of the eigenvalues} $0,\ 1,\ 2,\ ...$.

As in the proof of Theorem 4.1, we get the inequality of the Weyl symbols 
\begin{equation}
\label{ }
\sigma^\mathbf{Y}_\varpi(z^*,z)\ >  \sigma^\mathbf{N}_\varpi(z^*,z).
\end{equation} 
so that, by Theorem 3.2,
\begin{equation}
\label{ }
\mathbf{Y} > \mathbf{N}.
\end{equation}
The functional $Y(z^*,z)$ is a  polynomial with zero constant term. Therefore, its  anti-normal quantization $\mathbf{Y}$ is spanned by monomials $\widehat{\zeta^{*j}}\hat{z}^k, \ j+k>0$.
For the constant \emph{vacuum state} $1$
\begin{equation}
\label{eq:ann}
\widehat{z^*}1\ =\ \partial_{\zeta^*}1\ =\ 0,\quad \hat{z}1\ =\ (1^*\zeta)1\ =\ 0,
\end{equation}
so that  $H1=0$. This implies,  since $\mathbf{Y}$ is a nonnegative, that the vacuum state 1 is a fundamental  eigenstate  with zero eigenvalue.

The vacuum state  is  the a fundamental  eigenstate  with zero eigenvalue for the number operator $\mathbf{N}$. Moreover it is simple, i.e., the corresponding eigenspace of $\mathbf{N}$ is one-dimensional, and  the spectral gap of $N$ is 
 
 Then, by Glazman lemma ((see, e.g.,  \textsc{berezin-shubin}\cite[Appendix 1,Lemma 3.1]{Berezin}),  the operator inequality (\ref{eq:>}) implies that the the spectrum  of the operator $\mathbf{Y}$,, just as of $N$,  may contain only eigenvalues  in the open interval $-\infty<\lambda< 1$ and  the sum of their multiplicities  cannot be  greater than such sum for $N$  The latter sum is equal to 1. Since the interval $(-\infty,1)$ already contains  the zero spectral value of $\mathbf{Y}$, it follows that this spectral value is the  simple fundamental  eigenvalue. Thus $\mathbf{Y}$ has  a positive spectral gap.  \qed

\begin{remark}
By  \textsc{glassey-strauss}\cite{Glassey} the conformal invariance of Yang-Mills Lagrangian  implies  that  the energy-mass of a (global) solution to the Yang-Mills equations radiates out along the light cone. In particular,  its support in $\mathbb{R}^3$ spreads to infinity, i.e., classically,   no confinement  is possible. However, the Theorem implies that the Schroedinger equation  
 is globally solvable  over  any cutoff Gelfand-Kree triple, i.e.  there is a quantum confinement. 
 \end{remark}

\section{Appendix B}
\begin{proposition}
The global solutions of the Cauchy problem for modified classical Yang-Mills-Weyl  equations exist and are generated  generated by  global solutions with compactly supported Cauchy data.
\end{proposition} 
\proof (A sketch)
\begin{itemize}
  \item A solution $u(t,x)$ of the Cauchy  problem   for hyperbolic YMW evolution equations (\ref{eq:evolution})  at  $(t,x)\in\mathbb{M}$ is uniquely defined inside  the   open  cones $\{(t,y)\in\mathbb{M}:\  0\leq t<R ,\ |x-y|<R-t\}$ by the non-constrained initial data on the cones  bases at   $t=0$.
   \item   Any strip $\{(t,x)\in \mathbb{M}:\  0\leq t\leq T\}$ can be covered by such cones because
 the sizes of the smaller cubes and their positions  can be chosen arbitrarily.
   \item  Multiply an initial data $u(0,x)$ with an infinitely smooth cut-off function $\zeta(t,x), 0\leq \zeta(t,x)\leq 1$ supported by a ball $\mathbb{B}$ and equal to $1$ inside of a smaller ball $\mathbb{B}^\prime$ so that $u$ and $\zeta u$ are equal inside of the open cone over $\mathbb{B}^\prime$. 
  \item  Extend $\zeta u$   to $\mathbb{B}$. Presume that a solution  with cutoff  initial data exists for all  $t$.  
 \item Then the  uniqueness and global existence  of solutions of the  initial value problem on
  $\mathbf{R}^{1,3}$ follow from  the same  properties  of the   cutoff   initial data.

\item  Since the solutions of YMW evolution equations satisfy a non-linear wave equation (see \cite[Section 4]{Choquet} in the Lorentz gauge)  one gets, as in \textsc{\small eardley-moncrief}\cite{Eardley}, a non-linear  integral equation for larks with cutoff  initial data in a ball $\mathbb{B}$ by using the retarded (or advanced) fundamental solution of the linear wave equation perturbed  by non-linear  terms. Then, as in \cite{Eardley},  the  \emph{non-negative} energy-mass functionals $\mathcal{E}_{W}$  to verify the  existence of global solutions with cutoff  initial data (cp. \textsc{\small goganov,kapitanskii}\cite[Theorem 5]{Goganov}).Since the constraint equations are preserved by YMW evolution equations, it follows that  the set of smooth larks is generated  by  larks with  \emph{ cutoff  initial data}. \qed
 \end{itemize}


\begin{thebibliography}{20}
\bibitem{Agarwal} 
 Agarwal, C. S., and {W}olf, E., \textit{Calculus for functions of noncommuting operators and general phase-space methods in quantum mechanics}, Physical Rev. D \textbf{2}, 2161-2225, 1970. 

 \bibitem {Agranovich}
Agranovich M. S.,  \textit{ Elliptic Boundary Problems}, Partial differential operators. \textbf{IX} (Encyclopaedia of Mathematical Sciences), Springer, 1997.

 
\bibitem{Akhmedov}
Akhmedov, E., \textit{Neutrino physics}, ArXiv:  hep-ph/0001264.

\bibitem{Arnold}
 V. Arnold et al. (editors),
\textit{ Mathematics : Frontiers and Perspectives},
 American Mathematical Society, Providence,  2000.



\bibitem{Baez}
Baez, J.C., Segal, I.E.,and Zhou, Z.-F.: \textit{Introduction to algebraic and constructive
quantum field theory}, Princeton University Press, Princeton, 1992. 



\bibitem{Berezin-65}
Berezin, F.,  \textit{The Method of Second quantization}, Nauka. Moscow, 1965; Academic Press, 1966.

\bibitem{Berezin-71}
Berezin, F. A.,  \textit{Wick and antiWick symbols of operators},
 Math. USSR  Sb., \textbf{15}, 577-606,  1971.
 
\bibitem{Berezin-72}
Berezin, F. A.,  \textit{Covariant and  contravariant symbols of operators},
Mathematics of the USSR-Izvestiya, 6:5, 1117Ð1151, 1972.s

\bibitem{Berezin-91}
 Berezin, F. A., and Shubin, M. A.,  
 \textit{The Schroedinger equation}, Kluwer Academic Publishers, 1991.

\bibitem{Berline}
 Berline, N.,  Getzler, E., Vergne, M., \textit{Heat kernels and Dirac operators}, Springer-Verlag, 1992


 \bibitem{Bogoliubov}
Bogoliubov, N. N., and Shirkov, D. V.,  \textit{Introduction to the theory of quantized fields},  John 
Wiley, 1980.

\bibitem{Bogoliubov-90}
Bogoliubov, N. N.,  Logunov. A. A., Oksak, A. I., and Todorov, I. T.,
\textit{General Principles of Quantum Field Theory}, Kluwer, 1990.


 \bibitem{Choquet}
Choquet-Bruhat,Y., Christodoulou, D., \textit{Existence of global solutions of the Yang-Mills, Higggs and spinor Weyl equations in 3+1 dimensions}, Annales scientifiques  de l'E.N.S., 4e serie, 
\textbf{14}, 481-506, 1981. 

 \bibitem{Colombeau}
Colombeau, J.,   \textit{Differential Calculus and Holomorphy}, North-Holland Mathematics Studies,
 \textbf{84}, 1982.

\bibitem{Berezin} 
 Berezin, F. A., and Shubin, M. A.,  
 \textit{The Schr\"{o}dinger equation}, Kluwer Academic Publishers, 1991.

 \bibitem{Clay}
 \textit{Quantum Yang-Mills theory}, http://www.claymath.org/millenium-problems/yangmills-and-mass-gap

 \bibitem{Corrigan}
Corrigan, E.,  Ramond, P., \textit{A Note On The Quark Content Of Large Color Groups}, Phys. Lett.  \textbf{B87}, 73, 1979.

\bibitem{Dixmier}
Dixmier,J., \textit{Von Neumann algebras}, North-Holland,1981.

\bibitem{Doran} Doran, C., Lasenby A., \textit{Geometric Algebra for Physicists}, Cambridge University Press, 2003. 

\bibitem{Dynin-02}
Dynin, A.,  \textit{Feynman integral for functional Schroedinger equations}, Partial Differential Equations:M. Vishik Seminar (American Mathematical Society Translations-Series 2,\textbf{206}, Vishik, M.I, et al, eds), 65-80, 2002.


\bibitem{Durr}
D\"{u}rr, S., Fodor, Z., Frison, J., Hoebling, C.,  Hoffman, R.,  Katz, S. D.,  Krieg, S.,
Kurth, T., Lellouch, L., Lippert, T., Szabo, K. K., Vulvert, G., \textit{Ab-initio determination of light hadron masses}, Science, \textbf{322}, (2008),  no. 1224, 1224-1227.   arXiv:0906.3599 [hep-lat].

\bibitem{Dell'Antonio}Dell'Antonio, G. and Zwanziger, D., \textit{Every gauge orbit passes inside the Gribov horizon},  Comm. Math. Phys., \textbf{138} (1991), 259-299. 

\bibitem{Eardley}
Eardley, D., Moncrief, V,  \textit{The global existence of Yang-Mills-Higgs fields in 4-dimensional Minkowski space},
Comm. Math. Phys., \textbf{83} (1982), 193-212.

\bibitem{Fulton}
Fulton, {W}., Harris, J., \textit{Representation theory. A first course.}, Springer,1991


\bibitem{Faddeev}
Faddeev, L., Slavnov, A., \textit{Gauge fields, introduction to quantum theory}, Addison-{W}esley, 1991

\bibitem{Folland}
Folland, G., \textit{Harmonic analysis in phase space}, Princeton University press, 1989. 

\bibitem{Gelfand}
Gelfand, I. M., and Minlos, R. A., \textit{Solution of quantum field equations}, reprinted in Gelfand, I. M., \textit{Collected Papers}, V1, 462-465, Springer, 1987.


\bibitem{Gelfand-1}
Gelfand, I., Vilenkin, N..  \textit{Generalized functions},  \textbf{4}. Academic Press, 1964.

\bibitem{Glassey}
 Glassey, R. T.,  and Strauss, W. A , \textit{Decay of classical Yang-Mills fields}, \textbf{65}, no.1,
 Comm. Math. Phys. (1979), 1-13.
 
 \bibitem{Goganov}
Goganov, M. V., and Kapitanskii, L. V., \textit{Global solvability of the initial problem for Yang-Mills-Higs equations}, Zapiski LOMI,
\textbf{147} (1985), 18-48; J. Sov. Math.,   \textbf{37} (1987), 802-822.


\bibitem{Grothendieck}
Grothendieck, A., \textit{Rearrangements de fonctions et inegalites de convexite dans les algebres de von Neumann munies d'une trace}, Seminaire N. Bourbaki (1954-1956), Expo. no.113, 128-139.

\bibitem{Hatfield}
Hatfield, B., \textit{Quantum field theory of point particles and strings}, Addison-{W}esley, 1992.

\bibitem{Howe}
Howe, R., \textit{Quantum mechanics and partial differential equations}, J. Funct. Anal.,  \textbf{38} (1980),188Ð254.


\bibitem{Kree-1}
Kree, P., \textit{Calcule symbolique et second quantification}, C. R. Sc., Serie A, \textbf{284} (1978),  25-28.

\bibitem{Kree-2}
Kree, P., \textit{Methodes holomorphe et methodes nucleaires en analyse de dimension infinie et la theorie quantique des champs}, Lecture notes in mathematics, \textbf{644} (1978), 212-254.


\bibitem{Kronfeld}
Kronfeld, A. S., \textit{The weight of the {W}orld is Quantum Chromodynamics},  Science, \textbf{322}, (2008),  no. 5905, 1198-1199.
 
 \bibitem{Lang}
Lang, S., \textit{Differential and Riemannian manifolds}, Springer, 1995.

\bibitem{Lascar}
 Lascar, B.,  \textit{Une condition necessaire et suffisant 
d'ellipticite pour une class d'operateurs differentiells en dimension 
infinie}, Comm. Partial Diff. Equations, \textbf{2} (1977), 31-67.


 \bibitem{Lawson}
Lawson, H.B., Michelsohn, M-L.,  \textit{Spin Geometry}, Princeton University Press, 1980.


\bibitem{Narasimhan}
Narasimhan, M.S., Ramadas, T.R.,  \textit{Geometry of SU(2) gauge fields}
Communications in Mathematical Physics, \textbf{67} (1979), 121-136.

\bibitem{Ovchinnikov-1}
Ovchinnikov, V., \textit{s-numbers of mesurable operators}, Functional Analysis and Its Applications, 1970, 4:3, 236Ð242

\bibitem{Ovchinnikov-2}
Ovchinnikov, V., \textit{Compact operators relative to a von Neumann algebra},
Functional Analysis and Its Applications, 1972, 6:1, 31-34


\bibitem{Rempel}
Rempel, S., Schulze, B.-{W}., {Index theory of elliptic boundary problems}, Akademie-Verlag, 1982.

\bibitem{Kree-77}Kree, P., \textit{Calcule symbolique et second quantification des fonctions semiholomorphes},  CR Acad. Sc. Paris, \textbf{284A} (1974), 25--28. 


\bibitem{Reed}
Reed, M., Simon B.,
\textit{I. Methods of modern mathematical physics}, Academic Press, 1980

\bibitem{Simon}
Simon, B., \textit{Some Quantum Operators with Discrete Spectrum 
but Classically Continuous Spectrum}, Annals of physics, \textbf{146} (1983),  209-220. 


\bibitem{Schwarz}
Schwarz, G., \'{S}niatycki, J., \textit{Gauge symmetries of an extended phase spce for Yang-Mills and Dirac fields}, Annales de  l'Institut  H. Poincar\'{e} , Section A , \textbf{66}, no.1 (1997), 109-136.

\bibitem{Segal-60}
Segal, I., \textit{Quantization of nonlinear systems}, Journal of Mathematical Physics, \textbf{1}(1960), 468-488.

\bibitem{Segal-79}
Segal, I., \textit{The initial Problem for the Yang-Mills Equations}, Journal of  Functional Analysis, \textbf{33}(1979), 175-194. 

\bibitem{Shubin} Shubin, M.,\textit{Pseudodifferential operators and spectral theory}, Springer, 1987.

\bibitem{Singer}
Singer, I. M., \textit{Some remarks on the Gribov ambiguity}, Communications in Mathematical Physics,\textbf{60} (1978), 7-12.

\bibitem{Sniatycki}
 \'{S}niatycki, J., \textit{Regularity of constraint and reduction in the Minkowski space Yang-Mills-Dirac theory}, Annales de  l'Institut H. Poincar\'{e}., Section A , \textbf{70}, no.3 (1999), 277-293.
 
 
\bibitem{Strocchi}
Strocchi, F., S., \textit{Selected Topics of the General properties of Quantum Field Theory}, World Scientific, 1993.

 
 \bibitem{Takesaki}
Takesaki, M.,  \textit{Theory of Operator Algebras I}, Springer, 2001. 


 
 \bibitem{Weinberg}
 {W}einberg, S.,   \textit{The Quantum Theory of Fields}, Volume I, Cambridge University Press, 1995.

\bibitem{Wilczek}
Wilczek, F.,  \textit{Four Big Questions with Pretty Good Answers} in  
\textit{{W}erner Heisenberg Centennial Symposium "Developments in Modern Physics"} (Buschhorn, G.,  {W}ess,J., eds), 79-98, Springer, 2004.


\bibitem{Witten}
Witten, E., \textit{Physical law and the quest for mathematical understanding},
Bulletin of American Mathematical Society, \textbf{40}(2002), no.1, 21-29.


\end{thebibliography}
\end{document}